\documentclass[aps, prd, reprint,10pt, notitlepage, a4paper,floats, floatfix,amsmath, amssymb, amsfonts,superscriptaddress,showpacs, showkeys,nofootinbib,longbibliography]{revtex4-2}

\pdfoutput=1

\usepackage{epsfig}
\usepackage{graphics}
\usepackage{graphicx}
\usepackage{amsmath,amssymb,mathrsfs}
\usepackage{amsfonts}
\usepackage[usenames,dvipsnames]{xcolor}
\usepackage{wasysym}
\usepackage{times}
\usepackage{mathptmx}
\usepackage{gensymb}
\usepackage{appendix}
\usepackage{listings}
\usepackage{url}
\usepackage[normalem]{ulem}
\usepackage{alltt}
\usepackage[colorlinks]{hyperref}
\usepackage{cleveref}
\usepackage{longtable}
\usepackage{enumitem}
\setlist{nosep}
\usepackage{color}
\usepackage{calc}
\usepackage{bm}
\usepackage{times}
\usepackage{multirow}
\usepackage[varg]{txfonts}
\usepackage{float}
\usepackage{dcolumn}
\usepackage[nolist,nohyperlinks]{acronym}
\usepackage{xspace}
\usepackage[english]{babel}
\usepackage[abs]{overpic}
\usepackage{pict2e}
\usepackage[caption=false]{subfig}
\allowdisplaybreaks[1]
\usepackage[utf8]{inputenc}
\usepackage{gensymb}
\usepackage{bm}
\usepackage{stackengine}
\usepackage{boldline,multirow}
\usepackage{braket}
\usepackage{longtable}

\usepackage{tabularx}
\usepackage{rotating}

\DeclareMathAlphabet{\mathcalstd}{OMS}{cmsy}{m}{n}
\DeclareMathAlphabet{\mathpzc}{OT1}{pzc}{m}{it}

\newcommand{\AEI}{Max Planck Institute for Gravitational Physics (Albert Einstein Institute), Am M{\"u}hlenberg 1, Potsdam, 14476, Germany}
\newcommand{\Maryland}{Department of Physics, University of Maryland, College Park, MD 20742, USA}
\newcommand{\Nikhef}{Nikhef -- National Institute for Subatomic Physics, Science Park 105, 1098 XG Amsterdam, The Netherlands}
\newcommand{\UU}{Institute for Gravitational and Subatomic Physics (GRASP), \mbox{Utrecht University}, Princetonplein 1, 3584 CC Utrecht, The Netherlands}
\newcommand{\BarIlan}{Bar-Ilan University, Ramat Gan, 5290002, Israel}
\newcommand{\UP}{Institut f\"ur Physik und Astronomie, Universit\"at Potsdam, Haus 28, Karl-Liebknecht-Str. 24/25, 14476, Potsdam, Germany}
\newcommand{\bham}{School of Physics and Astronomy and Institute for Gravitational Wave Astronomy, University of Birmingham, Edgbaston, Birmingham, B15 2TT, United Kingdom}
\newcommand{\qmul}{Queen Mary University of London, Mile End Road, London, E1 4NS, UK}
\newcommand{\Cambridge}{Department of Applied Mathematics and Theoretical Physics, University of Cambridge, Wilberforce Road CB3 0WA Cambridge, United Kingdom}
\newcommand{\CIERA}{Center for Interdisciplinary Exploration and Research in Astrophysics (CIERA), Northwestern University, Evanston, IL 60201, USA}
\newcommand{\Liege}{Universit\'{e} de Li\`{e}ge, B-4000 Li\`{e}ge, Belgium}
\newcommand{\UIB}{IAC3–IEEC, Universitat de les Illes Balears, E-07122 Palma, Spain}
\newcommand{\PennState}{The Pennsylvania State University, University Park, PA 16802, USA}
\newcommand{\Rochester}{Center for Computational Relativity and Gravitation, Rochester Institute of Technology, Rochester, NY 14623, USA}
\newcommand{\Oregon}{University of Oregon, Eugene, OR 97403, USA}
\newcommand{\Adler}{Adler Planetarium, 1300 South DuSable Lake Shore Drive, Chicago, IL, 60605, USA}

\definecolor{RED}{HTML}{F5054F}
\definecolor{LIGHT_ORANGE}{HTML}{FDAA48}
\definecolor{DARK_ORANGE}{HTML}{FF5B00}
\definecolor{LIGHT_BLUE}{HTML}{448EE4}
\definecolor{DARK_BLUE}{HTML}{0343df}
\definecolor{LIGHT_GREEN}{HTML}{40C53C}
\definecolor{DARK_GREEN}{HTML}{02590F}

\hypersetup{citecolor=LIGHT_BLUE, urlcolor=DARK_BLUE, linkcolor=RED}

\newcommand{\SEOBHM}[0]{\textsc{SEOBHM}\xspace}
\newcommand{\SEOB}[0]{\textsc{SEOB}\xspace}
\newcommand{\SEOBNSBH}[0]{\textsc{SEOBNSBH}\xspace}
\newcommand{\SEOBT}[0]{\textsc{SEOBT}\xspace}
\newcommand{\PhenomXPHM}[0]{\textsc{PhenomXPHM}\xspace}
\newcommand{\PhenomXHM}[0]{\textsc{PhenomXHM}\xspace}
\newcommand{\PhenomXP}[0]{\textsc{PhenomXP}\xspace}
\newcommand{\XPNRTidal}[0]{\textsc{PhenomXPT}\xspace}

\def \msun  {\rm{M}_\odot}

\begin{document}

\title{Tests of general relativity with GW230529: \\ A neutron star merging with a lower mass-gap compact object}

\author{Elise M. S\"anger}
\email{elise.saenger@aei.mpg.de}
\affiliation{\AEI}

\author{Soumen Roy}
\email{soumen.roy@uclouvain.be}
\affiliation{\Nikhef}
\affiliation{\UU}

\author{Michalis Agathos}
\affiliation{\qmul}
\affiliation{\Cambridge}

\author{Ofek Birnholtz}
\affiliation{\BarIlan}

\author{Alessandra Buonanno}
\affiliation{\AEI}
\affiliation{\Maryland}

\author{Tim Dietrich}
\affiliation{\UP}
\affiliation{\AEI}

\author{Maria Haney}
\affiliation{\Nikhef}

\author{F\'elix-Louis Juli\'e}
\affiliation{\AEI}

\author{Geraint Pratten}
\affiliation{\bham}

\author{Jan Steinhoff}
\affiliation{\AEI}

\author{Chris Van Den Broeck}
\affiliation{\Nikhef}
\affiliation{\UU}

\author{Sylvia Biscoveanu}
\affiliation{\CIERA}

\author{Prasanta Char}
\affiliation{\Liege}

\author{Anna Heffernan}
\affiliation{\UIB}

\author{Prathamesh Joshi}
\affiliation{\PennState}

\author{Atul Kedia}
\affiliation{\Rochester}

\author{R. M. S. Schofield}
\affiliation{\Oregon}

\author{M. Trevor}
\affiliation{\Maryland}

\author{Michael Zevin}
\affiliation{\Adler}

\date{\today}

\begin{abstract}

On May 29, 2023, the LIGO Livingston observatory detected the gravitational-wave signal GW230529\_181500 from the merger of a neutron star with a lower mass-gap compact object. Its long inspiral signal provides a unique opportunity to test general relativity (GR) in a parameter space previously unexplored by strong-field tests. In this work, we performed parametrized inspiral tests of GR with GW230529\_181500. Specifically, we search for deviations in the frequency-domain GW phase by allowing for agnostic corrections to the post-Newtonian coefficients. We performed tests with the Flexible Theory Independent and Test Infrastructure For General Relativity frameworks using several quasicircular waveform models that capture different physical effects (higher modes, spins, tides).  We find that the signal is consistent with GR for all deviation parameters. Assuming the primary object is a black hole, we obtain particularly tight constraints on the dipole radiation at $-1$PN order of $|\delta\hat{\varphi}_{-2}| \lesssim 8 \times 10^{-5}$, which is a factor $\sim17$ times more stringent than previous bounds from the neutron star--black hole merger GW200115\_042309, as well as on the 0.5PN and 1PN deviation parameters. We discuss some challenges that arise when analyzing this signal, namely biases due to correlations with tidal effects and the degeneracy between the 0PN deviation parameter and the chirp mass. To illustrate the importance of GW230529\_181500 for tests of GR, we mapped the agnostic $-1$PN results to a class of Einstein-scalar-Gauss-Bonnet (ESGB) theories of gravity. We also conducted an analysis probing the specific phase deviation expected in ESGB theory and obtain an upper bound on the Gauss-Bonnet coupling of $\ell_{\rm GB} \lesssim 0.51~\msun$ ($\sqrt{\alpha_{\rm GB}} \lesssim 0.28$~km), which is better than any previously reported constraint.

\end{abstract}

\maketitle

\section{Introduction} \label{sec:intro}

On May 29, 2023, at 18:15:00 UTC, a gravitational wave (GW) signal emitted from the merger of two compact objects, most likely a black hole (BH) and a neutron star (NS)~\cite{LIGOScientific:2024elc}, was observed by the Advanced LIGO detector~\cite{KAGRA:2013rdx, LIGOScientific:2014pky}. This event was named GW230529\_181500 and will be referred to as GW230529 for brevity. It was observed only by the LIGO Livingston observatory, with a signal-to-noise ratio (SNR) of 11.6 and a false-alarm rate of less than 1 in 1000 years. Assuming that General Relativity (GR) is the correct theory to describe the GW230529 signal, the follow-up analyses estimated the source component masses to be $\ensuremath{3.6_{-1.1}^{+0.8}}~\msun$ and $\ensuremath{1.4_{-0.2}^{+0.6}}~\msun$ with a 90\% credible interval. This puts the primary squarely in the lower mass gap of $\sim 2 \textup{ -- } 5~\msun$, where few compact objects were observed~\cite{Fishbach:2020ryj, Farah:2021qom, LIGOScientific:2020zkf, LIGOScientific:2021duu, KAGRA:2021vkt}. This event, with the compact object within the hypothesized mass gap and with a long inspiral signal, provides an opportunity to test GR in a region of parameter space previously unexplored by strong-field tests of GR.

GW230529 is not the first observation to find support for compact objects in the lower mass gap. A number of recent studies that have found evidence for compact objects in the mass gap include observations of noninteracting binary systems~\cite{Thompson:2018ycv, Jayasinghe:2021uqb}, radio pulsar surveys~\cite{Barr:2024wwl}, and GWs from compact binary coalescences~\cite{LIGOScientific:2017vwq, LIGOScientific:2020aai, LIGOScientific:2020zkf, KAGRA:2021vkt}. When a binary is observed with component masses measured in the mass-gap range of $\sim 2 \textup{ -- } 5~\msun$, the immediate challenge is to determine whether the component in the gap is a NS or a BH. Such an identification would be remarkable because it would either increase the known maximum mass for a NS~\cite{Kalogera:1996ci, Rezzolla:2017aly} or decrease the known minimum mass for a BH from low-mass X-ray binary observations~\cite{Bailyn:1997xt, Farr:2010tu, Ozel:2010su}. However, it has not been conclusively proven whether
these components are BHs, NSs, or something else.

Recent theoretical research has suggested several potential exotic compact objects, such as gravastars~\cite{Mazur:2004fk}, boson stars~\cite{Liebling:2012fv}, or Planck-scale modifications of BH horizons~\cite{Lunin:2001jy, Cardoso:2016oxy}, which could also fall into this gap. Primordial BHs formed from overdense regions in the early Universe could also fill the lower mass gap~\cite{Carr:2016drx}. Besides astrophysical and primordial BHs in GR and exotic compact objects, NSs in alternative GR theories could also inhabit this gap. For example, axionic scalar-tensor theory with viable phenomenological equations-of-state (EOSs) can produce NSs with maximum masses larger than $2.5~\msun$, but below the $3~\msun$ threshold~\cite{Oikonomou:2024yzj}. Therefore, the GW230529 signal provides an opportunity to perform tests of GR to uncover new physics, as well as pointing to the existence of potential exotic compact objects.

General relativity is the simplest and most successful theory of gravity to date, and it has been rigorously validated by various experimental tests in our Solar System~\cite{Will:2014kxa}, observations of binary pulsars~\cite{Wex:2014nva, Shao:2017gwu}, cosmological data~\cite{Ferreira:2019xrr}, GWs~\cite{LIGOScientific:2016lio, LIGOScientific:2016dsl, LIGOScientific:2018dkp, LIGOScientific:2019fpa, LIGOScientific:2020tif, LIGOScientific:2021sio}, and massive BHs~\cite{Genzel:2024vou}. Recently, the evidence of a stochastic GW background with pulsar timing arrays~\cite{Bernardo:2023zna, Cannizzaro:2023mgc} has also led to some new gravitational tests. Despite its overall success, we do not know how to reconcile GR with quantum mechanics and how (or if) to employ GR to explain certain cosmological phenomena, such as dark matter and dark energy. These limitations have motivated the development of alternative theories of gravity, such as the Brans-Dicke~\cite{Brans:1961sx}, Horndeski scalar-tensor~\cite{Horndeski:1974wa}, Aether~\cite{Jacobson:2004ts}, Einstein-Gauss-Bonnet~\cite{Nojiri:2005vv, DeFelice:2010aj}, and Chern-Simons~\cite{Jackiw:2003pm, Alexander:2009tp} theories, and the effective-field-theory extension of GR~\cite{Endlich:2017tqa}, all of which have survived a wide range of experimental tests~\cite{Will:2014kxa, Yunes:2016jcc, Lyu:2022gdr, Takeda:2023wqn, Silva:2022srr}.

The observation of GW signals from coalescences of compact binary systems has opened the avenue for studying the two-body dynamics in strong  gravitational fields with high velocities approaching the speed of light. In contrast, other observations typically probe states of either weak gravitational fields---where spacetime curvature is negligible---or where velocities are well below the speed of light. However, in binary pulsars, the strongly gravitating bodies allow for strong-field tests of gravity, albeit for low velocities. Developing an accurate GW waveform model that captures the entire evolution of compact binaries is essential for determining source characteristics, validating the predictions of GR, and exploring potential alternative theories of gravity. However, the development of waveforms within specific alternative theories has not yet reached a level of maturity sufficient to perform model comparisons with GR. While numerical-relativity simulations in beyond-GR theories are becoming more prevalent, either by solving the full equations of motions~\cite{East:2020hgw, Bezares:2021dma, Figueras:2021abd, AresteSalo:2022hua, AresteSalo:2023mmd, AresteSalo:2023hcp, Corman:2022xqg, Corman:2024vlk} or through approximate treatments~\cite{Okounkova:2019dfo, Okounkova:2020rqw, Cayuso:2023aht}, there are not yet sufficient simulations in one specific theory to calibrate a semianalytic inspiral-merger-ringdown model in that theory. Therefore, semianalytical inspiral-merger-ringdown models in alternative theories of gravity that are compared to GW data either leave the merger like GR~\cite{Carson:2020iik, Carson:2020ter} or they introduce agnostic parameters in the merger that are then marginalized over during parameter estimation~\cite{Julie:2024fwy}. Another possibility is to directly compare numerical-relativity simulations in beyond-GR theories with observational data~\cite{Okounkova:2020rqw}, but those tests are limited by the lengths of the numerical waveforms and the region of parameter space where simulations are available. Nevertheless, there may be a more accurate alternative theory that we are not aware of.

The alternative models often propose variances in the post-Newtonian (PN) phase coefficients of the GW signal, which are derived solving perturbatively the two-body dynamics and gravitational radiation by a series expansion in the orbital velocity, intertwined with the strength of the gravitational field. This has led to the development of theory-agnostic approaches to parametrized tests of GR~\cite{Blanchet:1994ez,Arun:2006yw, Arun:2006hn, Yunes:2009ke, Mishra:2010tp}. The same framework can be used with search pipelines to detect non-GR signals~\cite{Narola:2022aob, Sharma:2023djw}. In these parametrized tests, an additional parameter is introduced into the PN coefficients to represent generic deviations from GR and a series of Bayesian parameter estimation analyses is performed for each of these parameters. The resulting posterior distribution of a deviation parameter indicates whether generic features in the GW signal suggest that GR is not the most probable theory of gravity. Importantly, claiming a GR-violation would require performing extensive tests that prove the GR deviation is not due to systematics due to waveforms' inaccuracy, lack of physical effects, environmental astrophysical effects, and gravitational lensing~\cite{Gupta:2024gun}. The parametrized test in the Bayesian framework has been extensively developed for the LIGO-Virgo-KAGRA (LVK) collaboration to test GR as part of the GW transient catalogs (GWTC)~\cite{LIGOScientific:2019fpa, LIGOScientific:2020tif, LIGOScientific:2021sio}. It has been performed using two implementations: the Flexible Theory-Independent (FTI) approach~\cite{LIGOScientific:2018dkp, Mehta:2022pcn}, which can be applied to any aligned-spin frequency-domain waveform model, and the Test Infrastructure For General Relativity (TIGER) approach~\cite{Li:2011cg, Agathos:2013upa, Meidam:2017dgf, Roy:2025gzv}, which is based on a (frequency-domain) phenomenological waveform family. In this study, we perform analyses of GW230529 using both approaches with the \textsc{SEOBNRv4}~\cite{Bohe:2016gbl, Cotesta:2018fcv, Cotesta:2020qhw, Dietrich:2019kaq, Matas:2020wab} and \textsc{IMRPhenomX}~\cite{Pratten:2020ceb, Garcia-Quiros:2020qpx, Colleoni:2023ple} families, respectively.

Since alternative theories often predict the functional form of the beyond-GR correction as a parametric variation of the PN terms, the measurements of phenomenological deviation parameters can be mapped onto specific alternative theories of gravity~\cite{Yunes:2016jcc, Berti:2018cxi, Tahura:2018zuq}. By using the leading order correction of an alternative gravity theory, we typically provide a theory-agnostic bound on the beyond-GR parameters predicted by that theory. In this work, we use this method to provide bounds on the Gauss-Bonnet coupling in the class of Einstein-scalar-Gauss-Bonnet (ESGB) theories of gravity with $f'(0)\neq 0$. We also perform a theory-specific test for ESGB by implementing all known corrections to the PN terms~\cite{Yagi:2011xp, Sennett:2016klh, Julie:2019sab, Shiralilou:2020gah, Shiralilou:2021mfl, Bernard:2022noq, Julie:2022huo, Julie:2022qux}. This includes newly computed corrections at 1.5PN order, as given in Appendix~\ref{ap:esgb-corrections}. The corrections to ESGB are larger for smaller BH masses, so GW230529 provides an excellent opportunity to probe relatively small values of the coupling. Previous tests for ESGB have placed constraints on the coupling of $\sqrt{\alpha_{\mathrm{GB}}} \lesssim 1.18$~km by combining bounds from multiple events~\cite{Lyu:2022gdr}, and another recent study of an eccentric binary obtained a bound of $\sqrt{\alpha_{\mathrm{GB}}} \lesssim 2.38$~km~\cite{Roy:2025xih}.

The rest of this paper is organized as follows. Section \ref{sec:model} explains the parametrized inspiral tests of GR used. Section \ref{sec:analysis} describes the setup of the analyses performed. The results and constraints on deviations from GR obtained for GW230529 are presented in Sec.~\ref{sec:results}. Finally, in Sec.~\ref{sec:esgb} we map the agnostic constraints to a specific modified gravity theory, namely ESGB gravity. In Appendix \ref{ap:priors}, we list the priors used in the analyses. Appendix \ref{ap:wrap-around} discusses possible false violations of GR when using too wide priors on the deviation parameters, which can cause wraparound in the waveform. Lastly, we list in Appendix \ref{ap:esgb-corrections} the ESGB corrections up to 1.5PN order in the GW phase. Throughout this paper we follow the convention of $G=c=1$.

\section{parametrized inspiral tests of GR} \label{sec:model}

In GR, the GW signal from the early inspiral of a quasicircular compact binary can be approximated using the PN formalism, which expands the waveform in powers of the velocity $v$, where $\mathcal{O}(v^{2n})$ relative to the leading order term is called the $n$PN order~\cite{Blanchet:2013haa,Buonanno:2009zt}. The frequency domain phase of a GW signal can then be obtained using the stationary phase approximation~\cite{Sathyaprakash:1991mt}. In GR, it is given by
\begin{equation}
    \label{waveformPhase}
    \begin{split}
        \Psi_{\ell m}^\text{GR}(f) = & 2\pi f t_c - \phi_c - \frac{\pi}{4} \\  & + \frac{3}{128\eta v^5}\frac{m}{2}  \sum_{n=0}^7\left( \psi_n^\text{GR} + \psi_{n(l)}^\text{GR} \log v \right) v^n
    \end{split}
\end{equation}
where $t_c, \phi_c$ are the time and phase at coalescence, $v = (2\pi fM/m)^{1/3}$ with $f$ the GW frequency, $M=m_1+m_2$ is the total mass, and $\eta=m_1 m_2/M^2$ is the symmetric mass ratio. Here, $\psi_n^\text{GR}$ and $\psi_{n(l)}^\text{GR}$ represent the $(n/2)$PN coefficients in GR, and they depend only on the intrinsic parameters of the binary. The subscript $(l)$ indicates the logarithmic terms that enter at 2.5PN and 3PN orders. It is also important to note that the 0.5PN term is absent in GR. The subscript $\ell m$ denotes the $(\ell,m)$ mode in the mode decomposition of the GW signal into spherical harmonics (not to be confused with the subscript $(l)$ denoting the coefficients of the $\log$ terms). The dominant mode is the $(2,2)$ mode, and other modes are referred to as higher modes.

In beyond-GR theories, these PN coefficients can be different from the ones in GR~\cite{Yunes:2016jcc, Berti:2018cxi, Tahura:2018zuq}. To test for deviations from GR during the inspiral, we therefore add a correction to the frequency domain phase of the form
\begin{equation}
    \delta \Psi_{\ell m}(f) = \frac{3}{128\eta v^5}\frac{m}{2} \left( \sum_{n=-2}^7 \delta\psi_n v^n + \sum_{n=5}^6 \delta\psi_{n(l)} v^n \log v \right), \label{eq:generic-phase-correction}
\end{equation}
where $\delta\psi_n$ and $\delta\psi_{n(l)}$ are the deviations in the $(n/2)$PN coefficients.

For parametrized inspiral tests, we introduce deviation parameters $\delta\hat{\varphi}_n$ and $\delta\hat{\varphi}_{n(l)}$ that are the fractional deviations of the corresponding PN coefficients in GR. We thus have that
\begin{align}
    \delta\psi_n &= \delta\hat{\varphi}_n \psi_n^\mathrm{GR}, \\
    \delta\psi_{n(l)} &= \delta\hat{\varphi}_{n(l)} \psi_{n(l)}^\mathrm{GR}.
\end{align}
When the $(n/2)$PN coefficient vanishes in GR (i.e., for $n=-2,1$), we instead let $\delta\psi_n = \delta\hat{\varphi}_n$ so that it is an absolute deviation normalized to the Newtonian coefficient. We do not test for $\delta\hat{\varphi}_5$ since this would give a constant phase shift and therefore be degenerate with the phase at coalescence. We also do not perform tests at $-0.5$PN. In this parametrization, GR is recovered in the limit $\delta\hat{\varphi}_{n}, \delta\hat{\varphi}_{n(l)} \rightarrow 0$.

We employ two different frameworks that can perform this type of test
of GR: FTI~\cite{Mehta:2022pcn} and TIGER~\cite{Agathos:2013upa, Meidam:2017dgf}. These two frameworks differ in the GR waveform models used and the exact implementation of the inspiral test. FTI uses the \textsc{SEOBNRv4} waveform family~\cite{Bohe:2016gbl, Cotesta:2018fcv, Cotesta:2020qhw, Dietrich:2019kaq, Matas:2020wab} and is only available for aligned spins, while TIGER uses the \textsc{IMRPhenomX} waveform family~\cite{Pratten:2020ceb, Garcia-Quiros:2020qpx, Colleoni:2023ple} and can use precessing spin waveforms. The main difference in the implementation occurs at the transition from a non-GR inspiral to the GR merger ringdown. In FTI, the testing GR corrections are added to the GW phase of an inspiral-merger-ringdown frequency-domain waveform from GR. The corrections to the frequency-domain phase are computed using Eq.~\eqref{eq:generic-phase-correction}, and then smoothly tapered off to zero using a windowing function so that the merger-ringdown signal coincides with the GR one (for details on how this is done, see Ref.~\cite{Mehta:2022pcn}). This is done for every mode separately and the corrections are then added together. In the TIGER framework, the parametrized deviations are incorporated into the GR phase coefficients of the \textsc{IMRPhenomXAS} model. This model represents the primary spherical harmonic radiation mode for the coalescence of nonprecessing binary black hole (BBH) systems. The inspiral phase of higher-order modes in the time domain is modeled as a scaling of the dominant quadrupole mode, allowing any deviations in the inspiral phase to propagate to the higher-order modes. In the phenomenological model, the separate inspiral and merger-ringdown segments are smoothly connected by ensuring $\mathcal{C}^1$ continuous condition in both phase and amplitude. When we introduce the variations in the phase coefficients, we simultaneously adjust the phase derivative to preserve the continuity. Deviations in the inspiral phase coefficients do not affect the postinspiral portion of the waveform. Although TIGER also has the option to introduce deviations in the postinspiral part of the waveform, we do not use that in this work.

\section{Analysis setup} \label{sec:analysis}

As baseline GR models for the FTI framework, we use the waveform models based on \textsc{SEOBNRv4HM\_ROM}~\cite{Cotesta:2020qhw}, which is the reduced order, frequency-domain version of the time-domain model \textsc{SEOBNRv4HM}~\cite{Bohe:2016gbl,Cotesta:2018fcv}. It is an effective-one-body waveform model for BBHs that assumes quasicircular orbits and aligned spins, and includes the modes: (2,2), (2,1), (3,3), (4,4), and (5,5)\footnote{We do not include the (5,5)-mode in our FTI analyses, but do use all other modes available.}. Its neutron star--black hole (NSBH) version \textsc{SEOBNRv4\_ROM\_NRTidalv2\_NSBH}~\cite{Dietrich:2019kaq, Matas:2020wab} contains only the dominant $(2,2)$-mode, and allows for tidal deformability of the secondary object and tidal disruption. We also use its binary neutron star (BNS) version \textsc{SEOBNRv4\_ROM\_NRTidalv2}, which has tidal effects on both components and contains the $(2,2)$ mode only. With the TIGER framework, we use the models based on \textsc{IMRPhenomXPHM}~\cite{Pratten:2020ceb, Garcia-Quiros:2020qpx}, which is a phenomenological waveform model that allows for precessing spins and includes the higher modes: (2,2), (2,1), (3,3), (3,2), and (4,4). Its BNS version \textsc{IMRPhenomXP\_NRTidalv2}~\cite{Dietrich:2019kaq, Colleoni:2023ple} allows for tides on both component objects and is $(2,2)$-mode only. To investigate the effect of waveform systematics due to the presence of precession, higher harmonics, or difference in waveform treatment, we also perform the parametrized test using \textsc{SEOBNRv4\_ROM}, \textsc{IMRPhenomXP}, and \textsc{IMRPhenomXHM} models. An overview of the models used and the physics they include is given in Table \ref{tab:waveforms}. The waveforms are generated using \textsc{Bilby TGR}~\cite{bilby-tgr} and \textsc{LALSimulation}~\cite{lalsuite, Wette:2020air}.

\begin{table*}
    \centering
    \begin{tabular}{llcccc}
        \hline
        \hline
        Waveform model & Short name & Color & Higher modes & Spin precession & Tides \\
        \hline
        \textsc{SEOBNRv4HM\_ROM}~\cite{Bohe:2016gbl, Cotesta:2018fcv, Cotesta:2020qhw} & \SEOBHM & {\color{LIGHT_ORANGE} $\bullet$} & \checkmark & - & - \\
        \textsc{SEOBNRv4\_ROM}~\cite{Bohe:2016gbl} & \SEOB &  & - & - & - \\
        \textsc{SEOBNRv4\_ROM\_NRTidalv2\_NSBH}~\cite{Dietrich:2019kaq, Matas:2020wab} & \SEOBNSBH & {\color{RED} $\bullet$} & - & - & \checkmark (NSBH) \\
        \textsc{SEOBNRv4\_ROM\_NRTidalv2}~\cite{Dietrich:2019kaq,Cotesta:2020qhw} & \SEOBT & {\color{DARK_ORANGE} $\bullet$} & - & - & \checkmark (BNS) \\
        \textsc{IMRPhenomXPHM}~\cite{Pratten:2020ceb} & \PhenomXPHM & {\color{LIGHT_BLUE} $\bullet$} & \checkmark & \checkmark & - \\
        \textsc{IMRPhenomXHM}~\cite{Garcia-Quiros:2020qpx} & \PhenomXHM &  & \checkmark & - & - \\
        \textsc{IMRPhenomXP}~\cite{Pratten:2020ceb} & \PhenomXP &  & - & \checkmark & - \\
        \textsc{IMRPhenomXP\_NRTidalv2}~\cite{Dietrich:2019kaq, Colleoni:2023ple} & \XPNRTidal & {\color{DARK_BLUE} $\bullet$} & - & \checkmark & \checkmark (BNS) \\
        \hline
        \hline
    \end{tabular}
    \caption{An overview of the different waveform models that we use in this work and the physics they include [higher modes or (2,2)-mode only, spin precession or aligned spins only, allowing for tidal effects]. We will refer to them throughout this paper using the shortened name and use the color indicated in the plots.}
    \label{tab:waveforms}
\end{table*}

For parameter estimation, we employ Bayesian inference using \textsc{Bilby}~\cite{Ashton:2018jfp, Romero-Shaw:2020owr, Smith:2019ucc}. We use the standard likelihood function assuming additive noise that is stationary and Gaussian. The one-dimensional posteriors for the deviation parameters are obtained by marginalizing over all other parameters. The sampling algorithm used is nested sampling with the \textsc{Dynesty} sampler~\cite{Speagle:2019ivv}. We only allow one deviation parameter to vary at a time and repeat the analysis for each deviation parameter. We use uniform priors for the deviation parameters. The priors used are listed in Appendix \ref{ap:priors}.

We analyze 128 seconds of data from the LIGO Livingston Observatory with a sampling frequency of 4096~Hz. The analysis is done over the frequency range 20~Hz to 1792~Hz when calculating the likelihood. The power spectral density used to describe the noise was produced with \textsc{BayesWave}~\cite{Cornish:2014kda, Littenberg:2014oda}.

\section{Results} \label{sec:results}

\begin{figure*}
    \centering
    \includegraphics[width=\textwidth]{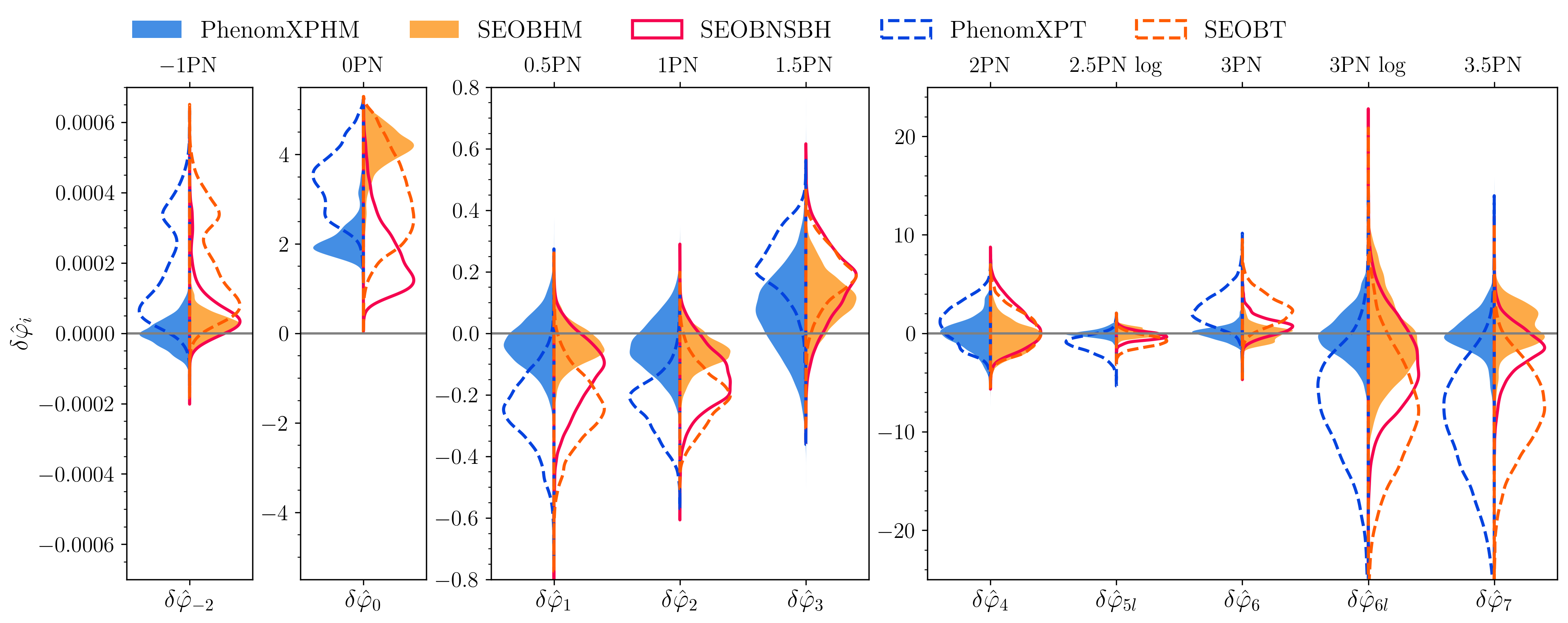}
    \caption{Posterior distributions for the different deviation parameters $\delta\hat{\varphi}_n, \delta\hat{\varphi}_{n(l)}$ for GW230529. The blue histograms are obtained with TIGER using the \textsc{IMRPhenomX} waveform family. The orange posteriors are results from FTI using the \textsc{SEOBNRv4} waveform family. The filled violins are for BBH waveforms models, while the dashed lines are BNS models that include tidal effects on both components, and the red solid line is an NSBH model with tides only on the secondary. GR is recovered at $\delta\hat{\varphi}_i=0$, which is indicated by the horizontal gray line. All results are consistent with GR except for 0PN, which will be discussed in more detail in Sec. \ref{sec:0pn}.}
    \label{fig:violin-plot}
\end{figure*}

\begin{table*}
    \centering
    \begin{tabular}{lcrcrcrcrcr}
        \hline
        \hline
        \multirow{2}*{Parameter} & \multicolumn{2}{c}{\SEOBHM} & \multicolumn{2}{c}{\SEOBNSBH} & \multicolumn{2}{c}{\PhenomXPHM} & \multicolumn{2}{c}{\XPNRTidal} \\
         & $\delta\hat{\varphi}_i$ & $Q_{\mathrm{GR}}$ & $\delta\hat{\varphi}_n$ & $Q_{\mathrm{GR}}$ & $\delta\hat{\varphi}_i$ & $Q_{\mathrm{GR}}$ & $\delta\hat{\varphi}_i$ & $Q_{\mathrm{GR}}$ \\
        \hline
        $\delta\hat{\varphi}_{-2}$ & $2.5_{-6.1}^{+7.2} \times 10^{-5}$ & 23\% & $4.6_{-6.8}^{+22} \times 10^{-5}$ & 14\% & $0.1_{-6.8}^{+8.7} \times 10^{-5}$ & 49\% & $14_{-12}^{+27} \times 10^{-5}$ & 3\% \\
        $\delta\hat{\varphi}_{0}$ & $4.2_{-1.2}^{+0.6}$ & 0\% & $1.6_{-0.8}^{+1.7}$ & 0\% & $2.0_{-0.4}^{+1.2}$ & 0\% & $3.4_{-1.0}^{+1.1}$ & 0\% \\
        $\delta\hat{\varphi}_{1}$ & $-0.04_{-0.10}^{+0.11}$ & 75\% & $-0.12_{-0.21}^{+0.15}$ & 90\% & $-0.04_{-0.15}^{+0.15}$ & 67\% & $-0.25_{-0.22}^{+0.17}$ & 99\% \\
        $\delta\hat{\varphi}_{2}$ & $-0.06_{-0.11}^{+0.11}$ & 82\% & $-0.14_{-0.15}^{+0.16}$ & 93\% & $-0.04_{-0.14}^{+0.15}$ & 69\% & $-0.23_{-0.16}^{+0.13}$ & 99\% \\
        $\delta\hat{\varphi}_{3}$ & $0.11_{-0.16}^{+0.15}$ & 13\% & $0.17_{-0.21}^{+0.18}$ & 9\% & $0.07_{-0.23}^{+0.21}$ & 30\% & $0.23_{-0.17}^{+0.15}$ & 2\% \\
        $\delta\hat{\varphi}_{4}$ & $-0.4_{-2.3}^{+2.7}$ & 61\% & $0.5_{-2.6}^{+3.7}$ & 40\% & $-0.1_{-2.9}^{+2.8}$ & 51\% & $1.0_{-3.0}^{+3.0}$ & 32\% \\
        $\delta\hat{\varphi}_{5l}$ & $0.01_{-0.60}^{+0.79}$ & 49\% & $-0.27_{-0.68}^{+0.74}$ & 73\% & $-0.03_{-0.68}^{+0.81}$ & 54\% & $-1.1_{-1.7}^{+1.2}$ & 94\% \\
        $\delta\hat{\varphi}_{6}$ & $0.0_{-1.3}^{+1.6}$ & 49\% & $0.7_{-1.4}^{+1.8}$ & 21\% & $0.1_{-1.2}^{+1.4}$ & 46\% & $2.4_{-2.1}^{+2.9}$ & 3\% \\
        $\delta\hat{\varphi}_{6l}$ & $-0.9_{-6.1}^{+8.1}$ & 59\% & $-3.0_{-6.7}^{+7.3}$ & 75\% & $-0.2_{-4.0}^{+4.5}$ & 54\% & $-7.0_{-9.9}^{+7.2}$ & 95\% \\
        $\delta\hat{\varphi}_{7}$ & $0.6_{-4.3}^{+3.3}$ & 40\% & $-1.5_{-4.7}^{+3.4}$ & 77\% & $-0.1_{-4.7}^{+3.7}$ & 52\% & $-8.9_{-10}^{+7.7}$ & 97\% \\
        \hline
        \hline
    \end{tabular}
    \caption{The values for the different deviation parameters $\delta\hat{\varphi}_n, \delta\hat{\varphi}_{n(l)}$ obtained with GW230529. Displayed are the median values and 90\% confidence intervals inferred. We also display the quantile $Q_{\mathrm{GR}}$ corresponding to the GR value $\delta\hat{\varphi}_i=0$.}
    \label{tab:bounds}
\end{table*}

The posteriors for the different deviation parameters are shown in Fig.~\ref{fig:violin-plot}. The blue violins on the left are obtained with TIGER, while the orange violins on the right are results from FTI. The filled violins do not have tidal effects, while the unfilled violins with dash and continuous lines are for waveform models including tidal effects. Table \ref{tab:bounds} shows the values for the deviation parameters found for GW230529 and the quantile corresponding to GR. We see that all results obtained are consistent with GR, except for the 0PN results. We will first discuss in Sec.~\ref{sec:bbhs} in more detail the results obtained with BBH waveforms. Then, in Sec.~\ref{sec:tides}, we explain the
effects that tides in the NSBH and BNS models have on the results. Finally, we will discuss the 0PN results in more detail in Sec.~\ref{sec:0pn}.

\subsection{Binary black hole waveforms} \label{sec:bbhs}

The results for the aligned-spin \PhenomXHM, spin-precessing (2,2)-mode \PhenomXP, and aligned-spin (2,2)-mode \SEOB are not shown in the violin plot, but they are similar to the spin-precessing \PhenomXPHM and aligned-spin \SEOBHM results, respectively. It is not surprising that the results for spin-precessing \PhenomXPHM and aligned-spin \PhenomXHM agree, since for GW230529 there was no evidence found for spin-precession (the posteriors on the effective precessing spin were uninformative~\cite{LIGOScientific:2024elc}). It was also expected that the results for \PhenomXPHM and \PhenomXP agree well, because for the SNR and source parameters of GW230529, the higher modes are not expected to strongly contribute. The posteriors for \SEOB are slightly wider than for \SEOBHM, but otherwise agree well. The increase in width is most likely due to a slight loss in SNR when not including the higher-order modes.

Let us compare the FTI and TIGER results for BBH waveform models by comparing \PhenomXPHM (blue filled) and \SEOBHM (orange filled) results with each other. Differences between these results can come from waveform systematics, for example the details of how the waveforms are constructed and the approximations made therein, or from differences in the implementation of the inspiral test of GR in FTI and TIGER. This is a good consistency check to see that both pipelines work as intended. Looking at the filled violins in Fig.~\ref{fig:violin-plot}, we see that the results are consistent with each other (except for $\delta\hat{\varphi}_0$). There are sometimes some minor differences between the \SEOBHM and \PhenomXHM, e.g., one posterior is slightly wider than the other, but there are no major differences.

\subsection{Waveforms with tidal effects} \label{sec:tides}

When looking at the waveforms with tidal effects in Fig.~\ref{fig:violin-plot}, we see that there are no major differences between \XPNRTidal (blue dashed line) and \SEOBT (orange dashed line). However, we notice that there are significant differences compared to the waveforms without tidal effects (filled violins). The posteriors for \XPNRTidal and \SEOBT, which have tides on both compact objects, are shifted away from GR and are wider compared to the results without tidal effects. This also happens to a lesser amount for \SEOBNSBH (red solid line), which only has tides on the secondary object.

The \textsc{NRTidalv2} models \cite{Dietrich:2019kaq} modify the BBH waveforms to account for differences in the waveform coming from tides and other NS matter effects. Both the phase and amplitude are modified due to tidal effects, which are modeled using the tidal deformabilities $\Lambda_i$ of the two objects. For BHs, the tidal deformability is zero, so the NSBH model \SEOBNSBH enforces $\Lambda_1=0$. We do not expect changes in the amplitude to significantly influence our results because, in the parametrized inspiral tests, we only modify the frequency domain phase. We therefore only expect the change in phase evolution to influence the posteriors on the deviation parameters for the BNS and NSBH models.

The merger-ringdown is modeled differently for BNS and NSBH models. The merger frequency of the \XPNRTidal and \SEOBT models is calibrated to NR. Above this frequency, the amplitude is smoothly tapered off to zero in an agnostic way~\cite{Dietrich:2019kaq}. The reason for this is that the merger-ringdown of BNSs is not modeled well enough, and, more importantly, the merger frequency is typically high enough that the postmerger signal is outside the sensitive frequency band of current detectors, hence, no bias is expected during parameter estimation~\cite{Dudi:2018jzn}. When using these models for GW230529, the merger frequency is lower and inside the frequency band analyzed. To check that the amplitude tapering is not influencing our results, we also did an analysis with a maximum frequency of 400~Hz, which is below the merger frequency. No differences were found between the posteriors, so the amplitude tapering is not causing the shift away form GR that we observe when using \XPNRTidal and \SEOBT.

For \SEOBNSBH, the merger-ringdown is modeled differently. For NSBHs, tidal disruption of the NS can happen when the tidal disruption frequency is lower than the ringdown frequency~\cite{Matas:2020wab}. If this happens, then the amplitude is smoothly tapered off above the tidal disruption frequency. Otherwise, the merger-ringdown is modeled similarly to the merger-ringdown of BBHs, with a slightly suppressed amplitude. For GW230529, the NS most likely gets tidally disrupted~\cite{LIGOScientific:2024elc, Chandra:2024ila}. However, the tidal disruption frequency is above the maximum frequency used in our analysis~\cite{Matas:2020wab, LIGOScientific:2024elc}. We therefore do not expect tidal disruption to influence our results.

During inspiral, both the phase and amplitude are modified due to tidal effects. Because the tidal deformability is always positive (at least for NSs), the change in phase is always in the same direction. The deviation parameters try to cancel this effect by changing the phase in the opposite way. This means that a nonzero $\Lambda$ will always lead to a specific $\delta\hat{\varphi}_n$ to shift toward either positive or negative values (depending on the sign of the PN coefficient in GR), but not both. This explains why the deviation parameters all shift away from zero for the BNS and NSBH models compared to the BBH models. The shift is less strong for the NSBH model since there is only one tidal parameter that is nonzero so the tides have less of an effect. The extra tidal parameters also lead to broadening of the posteriors, especially for low SNR signals where their values are not well measured. Of course, this explanation is oversimplified due to correlations between the tides and other GR parameters as well as the deviation parameters, but nonetheless it explains why there is a shift away from GR and not just a broadening of the posteriors.

\begin{figure}
    \raggedright
    \includegraphics[width=0.85\columnwidth]{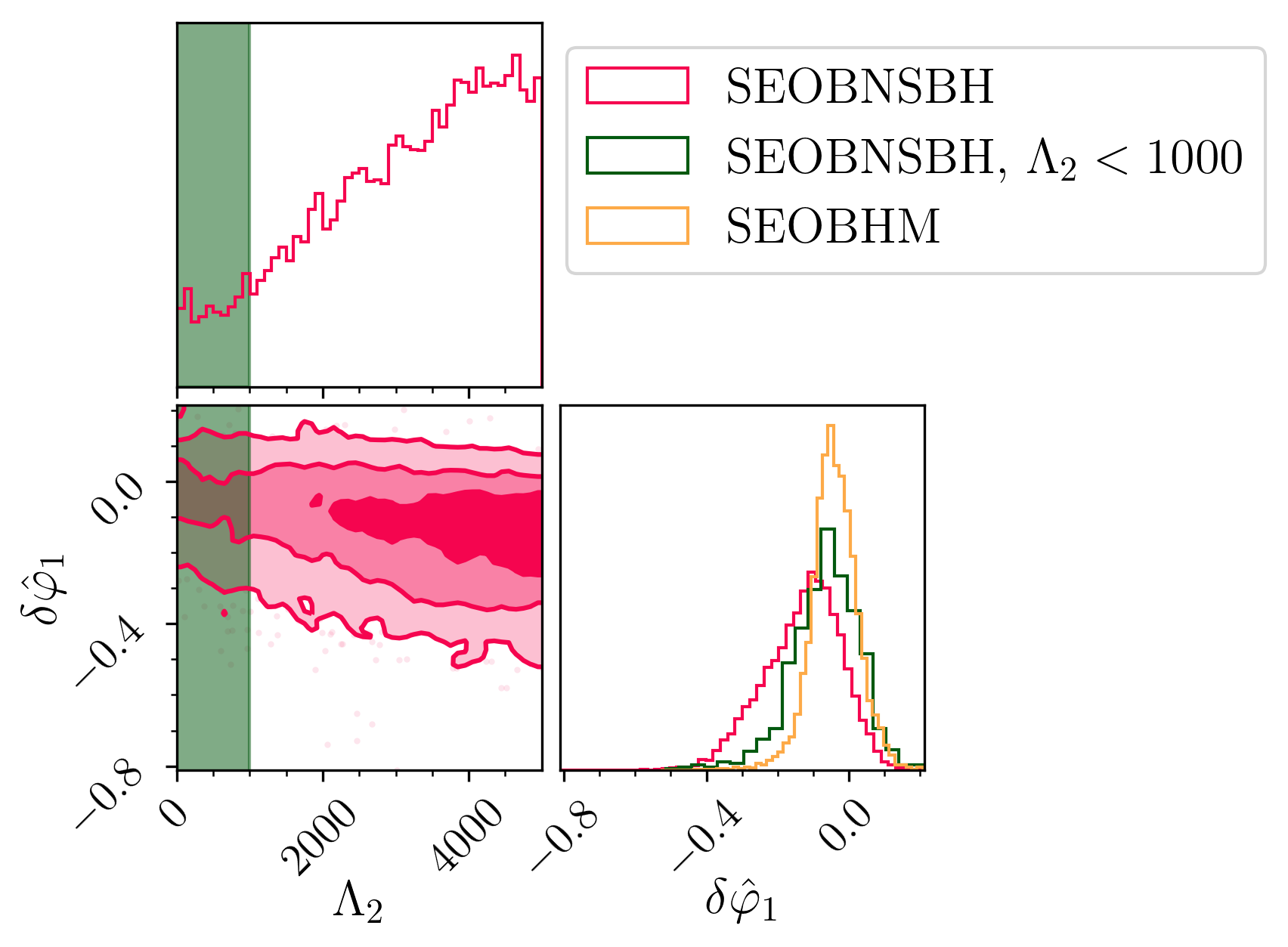}
    \includegraphics[width=\columnwidth]{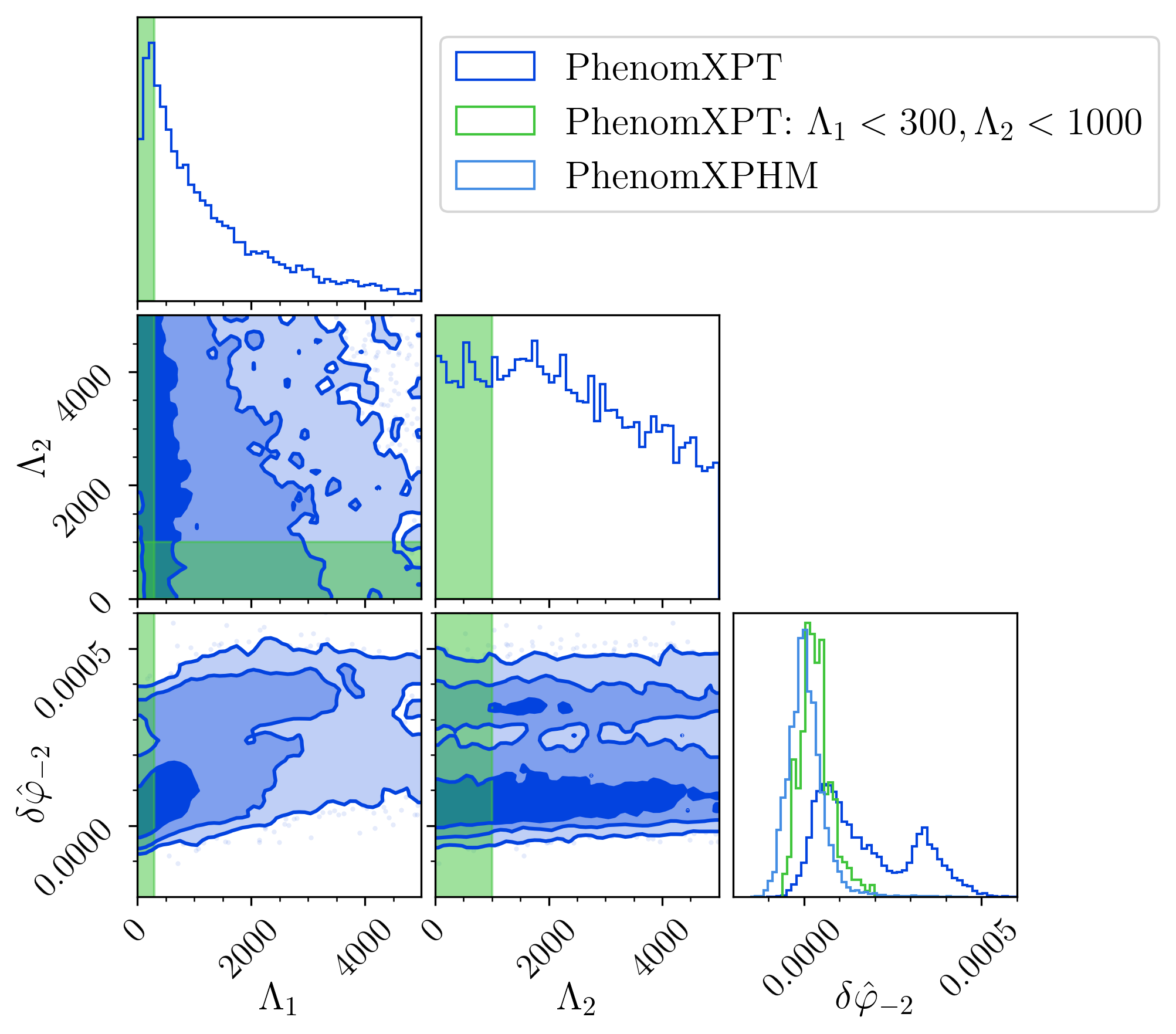}
    \includegraphics[width=\columnwidth]{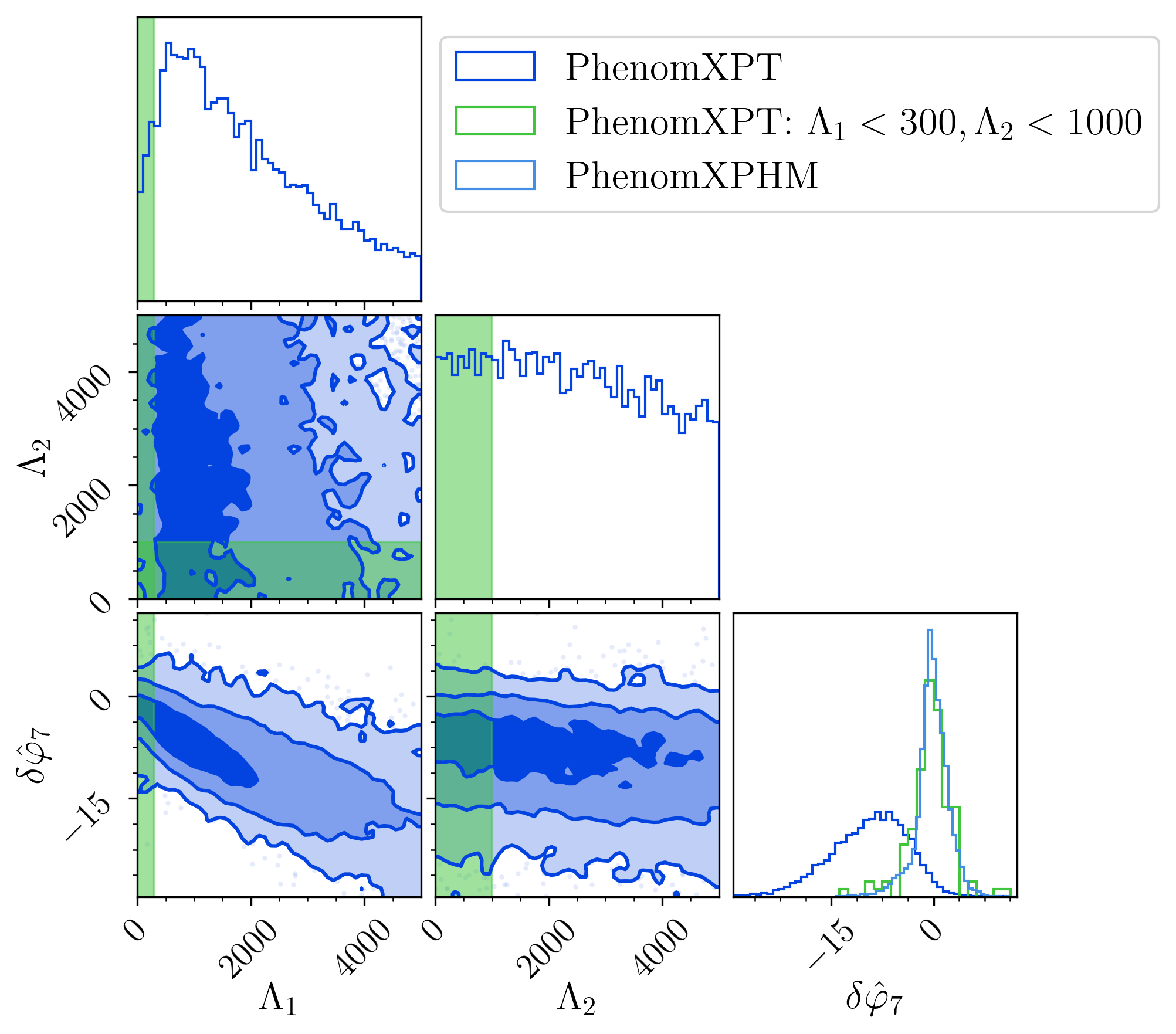}
    \caption{2D posteriors between the tidal deformability $\Lambda_i$ and deviation parameters $\delta\hat{\varphi}_n$ (red and dark blue). The green shaded regions are more realistic ranges for the tides, which, when restricting the posterior to those regions, give the green histograms. The orange and light blue histograms are the posterior obtained with the comparable BBH models. The top plot is for an NSBH waveform, while the other two are with a BNS waveform. We see that there is a correlation between the tides and the deviation parameters, which leads to a shift away from GR. When restricting the tides to more reasonable values, this shift disappears.}
    \label{fig:tidal-effect}
\end{figure}

To check that this reasoning is correct, we look at the $\Lambda_i$ -- $\delta\hat{\varphi}_n$ posteriors in Fig.~\ref{fig:tidal-effect}. The top plot shows the posteriors from an NSBH run for the 0.5PN deviation parameter (red). We see in the 2D posterior that $\delta\hat{\varphi}_1$ shifts away from 0 for larger values of $\Lambda_2$. This can be seen in the histogram where it has more support for negative values of $\delta\hat{\varphi}_1$ than is the case for the corresponding BBH run (orange). If we restrict the tidal deformability to $\Lambda_2 < 1000$ (green shaded area), which is a more realistic range based on the mass of the secondary and current constraints on the NS EOS~\cite{Huth:2021bsp} and it is consistent with what was found using an astrophysically motivated prior in Ref.~\cite{LIGOScientific:2024elc}, then the shift away from GR disappears and the posterior (green histogram) becomes similar to the BBH one.

In the middle and bottom plots of Fig.~\ref{fig:tidal-effect} we look at BNS runs (dark blue) for the $-1$PN deviation parameter (middle) and 3.5PN deviation parameter (bottom). Here we see the same effect where $\delta\hat{\varphi}_{-2}$ and $\delta\hat{\varphi}_{7}$ are correlated with $\Lambda_1$ and they shift away from GR for large values of $\Lambda_1$. If we now restrict the tides to small values of $\Lambda_1 < 300$\footnote{More realistically, using the EOS set of~\cite{Huth:2021bsp}, the tidal deformability for a NS with masses above $2.5 \msun$ (if such masses are supported) would be $\Lambda \lesssim 5$. Restricting $\Lambda_1$ to such low values would however leave us with too few samples to say anything sensible. We therefore went with a larger cutoff value that is a balance between the number of samples left and keeping the tides realistic.} and the more realistic values of $\Lambda_2 < 1000$ (green shaded area)~\cite{Huth:2021bsp, LIGOScientific:2024elc}, we once again see that this leads to the posterior (green histogram) shifting to the BBH one (light blue). This behavior is also seen for the other deviation parameters.

We conclude that the shift away from GR of the posteriors on the deviation parameters for the \SEOBNSBH and especially the \XPNRTidal and \SEOBT waveform models is most likely due to correlations between the deviation parameters and the tidal deformability combined with the latter not being well constrained. Limiting the tidal deformability of the two objects to more realistic values based on their masses removes this shift away from GR. Since the NSBH and BNS results with the tides constrained to realistic values are similar to the BBH results, we will use the BBH results from now on (unless otherwise specified), even though GW230529 is most likely a NSBH~\cite{LIGOScientific:2024elc}.

\subsection{Degeneracy between 0PN-deviation parameter and chirp mass} \label{sec:0pn}

At low frequencies (i.e., during early inspiral), the phase of a GR waveform is dominated by the 0PN term. In our testing GR waveform model, the phase contribution from the 0PN term is given by
\begin{equation}
    \Psi_{\ell m}^{0\mathrm{PN}} (f) = \frac{3}{128\eta v^5} \frac{m}{2} \left(1+\delta\hat{\varphi}_0\right)\psi_0^{\mathrm{GR}} = C_0 \frac{1+\delta\hat{\varphi}_0}{\mathcal{M}_c^{5/3}} f^{-5/3},
\end{equation}
where $C_0$ is a constant that does not depend on the intrinsic parameters of the binary and $\mathcal{M}_c = \eta^{3/5} M$ is the chirp mass. This means that in the frequency regime where the 0PN term is dominant, the deviation parameter $\delta\hat{\varphi}_0$ is degenerate with the chirp mass of the binary, so the two are strongly correlated (see also Sec.~IVC in Ref.~\cite{Mehta:2022pcn}). Assuming that the binary is consistent with GR and that the true chirp mass is equal to $\mathcal{M}_c^{\mathrm{GR}}$, this correlation is given by
\begin{equation}
    \delta\hat{\varphi}_0 = \left( \frac{\mathcal{M}_c}{\mathcal{M}_c^{\mathrm{GR}}} \right)^{5/3} -1. \label{eq:0pn-degeneracy}
\end{equation}

For low chirp-mass systems, the late inspiral and merger-ringdown are outside the frequency band observed by the GW detectors. For such systems, this degeneracy at 0PN means that, when doing inspiral tests of GR at 0PN, the chirp mass can take any value and $\delta\hat{\varphi}_0$ can be used to compensate without significantly altering the waveform. For high enough chirp mass, the late inspiral and merger-ringdown move into the frequency band used for analysis. This eventually breaks the degeneracy, and such high values of the chirp mass would then be disfavored. This means that the $\mathcal{M}_c$ -- $\delta\hat{\varphi}_0$ posterior is more or less free to move along the line of the degeneracy between the two, until it reaches too high values of the chirp mass.

\begin{figure}
    \centering
    \includegraphics[width=\columnwidth]{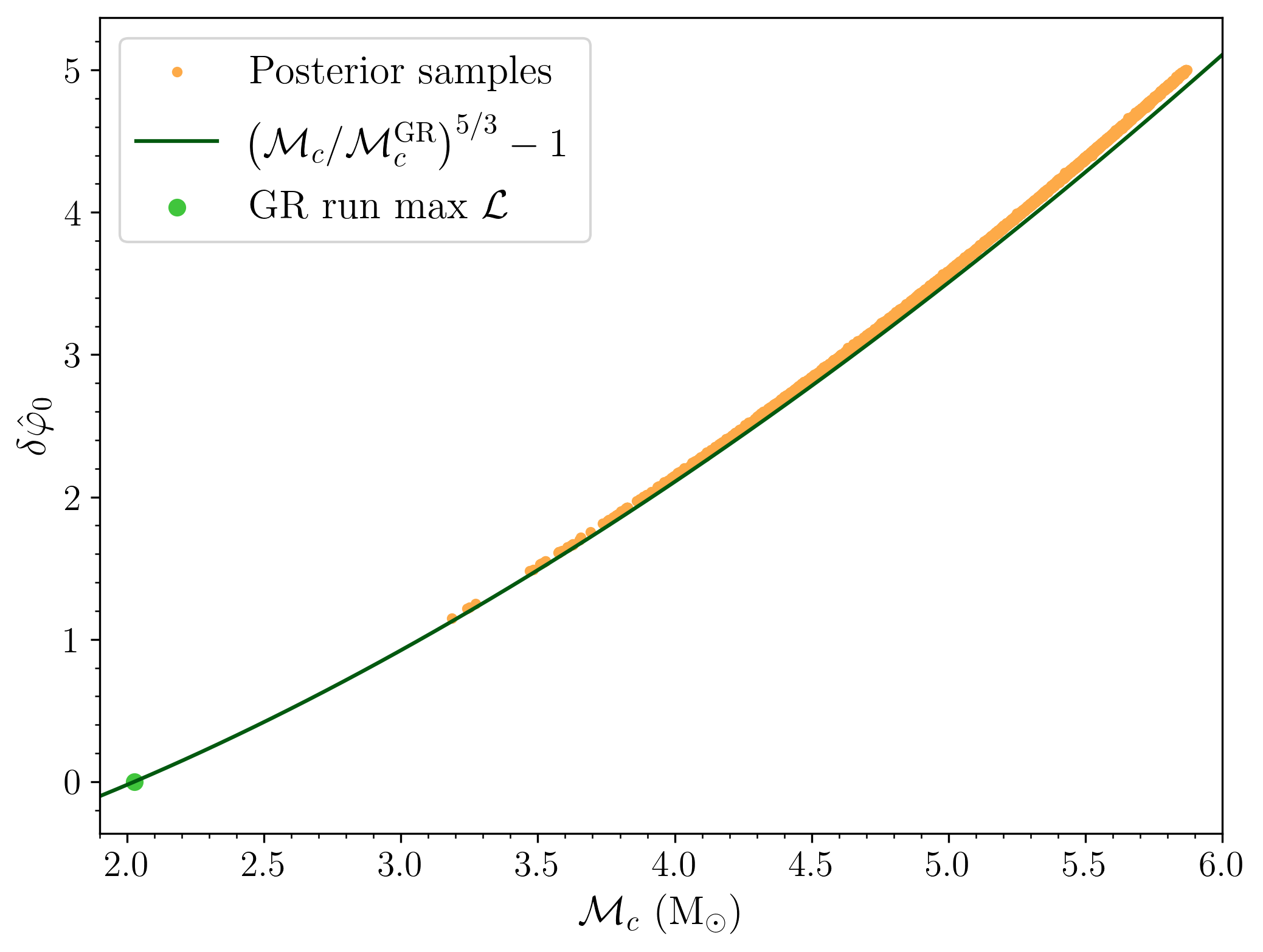}
    \caption{2D posterior between the chirp mass $\mathcal{M}_c$ and the 0PN deviation parameter $\delta\hat{\varphi}_0$ using the \SEOBHM waveform model (orange). The green dot indicates the maximum likelihood values obtained with a GR run. The dark green line is the expected degeneracy between $\mathcal{M}_c$ and $\delta\hat{\varphi}_0$ based on the GR maximum likelihood chirp mass $\mathcal{M}_c^\mathrm{GR}$ as per Eq.~\eqref{eq:0pn-degeneracy}. We see that the posterior follows the expected correlation.}
    \label{fig:dchi0-mc}
\end{figure}

Figure \ref{fig:dchi0-mc} shows the 2D posterior of the chirp mass $\mathcal{M}_c$ and 0PN deviation parameter $\delta\hat{\varphi}_0$ for GW230529 obtained with \SEOBHM (orange). The maximum likelihood sample from a GR run with \SEOBHM is highlighted in green. This is also the value used in Eq.~\eqref{eq:0pn-degeneracy} to compute the correlation (dark green). We see that the posterior closely follows the correlation that is expected from the 0PN degeneracy.

\begin{figure}
    \centering
    \includegraphics[width=\columnwidth]{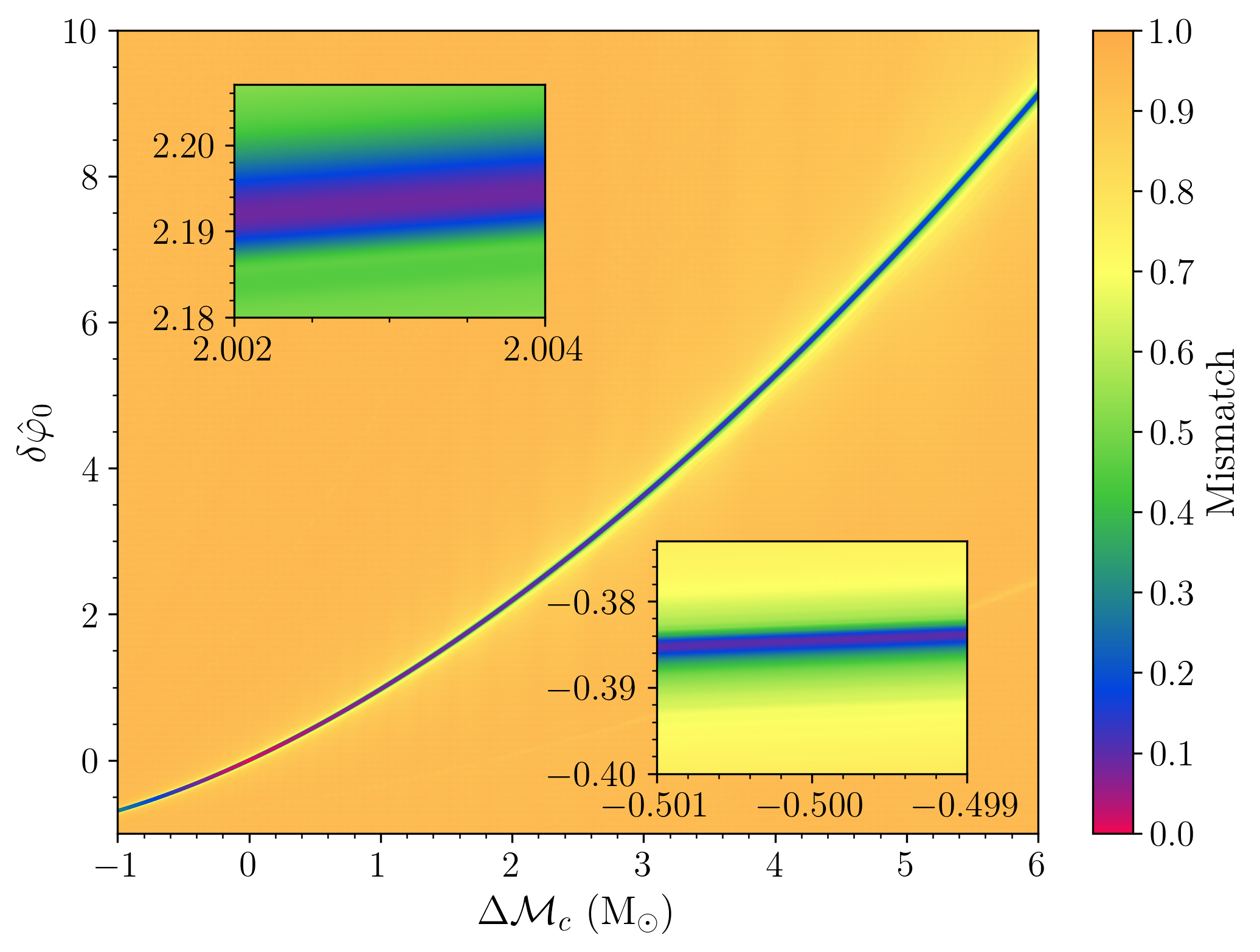}
    \caption{Mismatch between the maximum likelihood waveform from a GR run using \PhenomXPHM and non-GR waveforms as a function of chirp mass $\mathcal{M}_c$ and the 0PN deviation parameter $\delta\hat{\varphi}_0$. We clearly see the presence of the degeneracy from Eq.~\eqref{eq:0pn-degeneracy}. The insets show a zoom-in on two regions around the degeneracy line. We notice that the low mismatch region becomes wider for larger chirp mass.}
    \label{fig:mismatch}
\end{figure}

To check that this degeneracy should be present in GW230529-like signals, we compute the mismatch between the maximum likelihood GR waveform and non-GR waveforms using \PhenomXPHM. We vary the chirp mass and 0PN deviation coefficient, but keep all other parameters the same, and the mismatch is minimized over time and phase. Figure \ref{fig:mismatch} gives a map of these mismatches. We see that there is a line along which the mismatch is small. This line corresponds to the correlation in Eq.~\eqref{eq:0pn-degeneracy}. We notice that the low mismatch region becomes wider for larger chirp mass (see also the insets).

\begin{figure}
    \centering
    \includegraphics[width=\columnwidth]{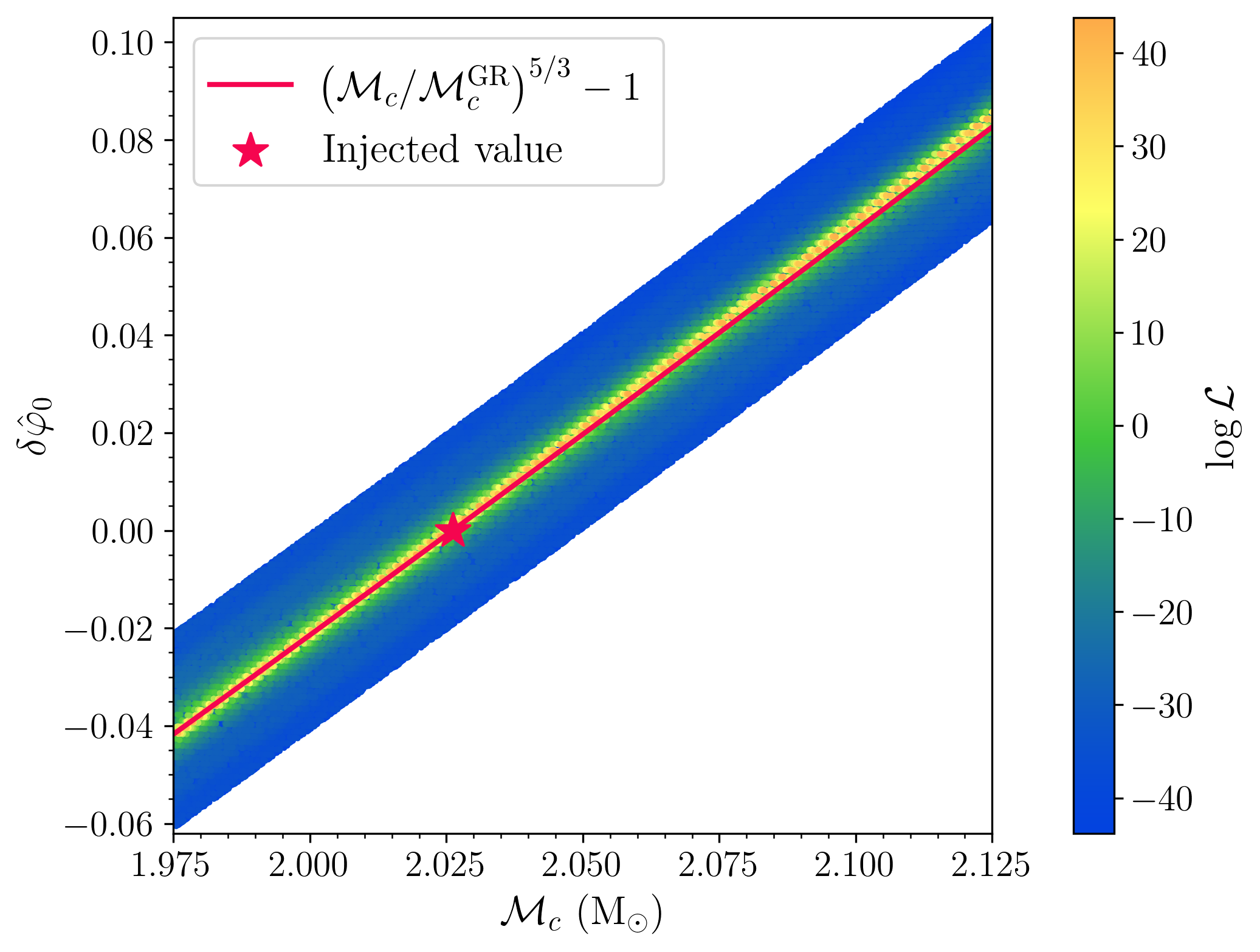}
    \caption{Likelihood for non-GR waveforms as a function of chirp mass $\mathcal{M}_c$ and the 0PN deviation parameter $\delta\hat{\varphi}_0$ when injecting the maximum likelihood waveform from a GR run using \SEOBHM. The injected values are indicated by the red star, and the red line is the expected degeneracy according to Eq.~\eqref{eq:0pn-degeneracy}. We see that the high likelihood region is sharply peaked and follows the expected correlation.}
    \label{fig:logl}
\end{figure}

As an extra check that this strong correlation should indeed be present in a GR signal with similar source parameters as GW230925, we inject the maximum likelihood GR waveform using \SEOBHM into a zero-noise realization of the LIGO Livingston detector with the same power spectral density as at the time of GW230529. We then evaluate the likelihood for random values of the chirp mass and 0PN deviation parameter, while keeping the other parameters the same as for the injected waveform. This gives the likelihood map shown in Fig.~\ref{fig:logl}. The red star indicates the injected values, and the red line shows the expected degeneracy in the waveform according to Eq.~\eqref{eq:0pn-degeneracy}. We see that the maximum likelihood region (yellow colors) indeed follows this line. There is no visible gradient along the line, which confirms that the waveform does not significantly change along the line (at least not within this chirp mass range).

When looking at Fig.~\ref{fig:dchi0-mc}, we also notice that the posterior is away from the GR value of $\delta\hat{\varphi}_0 = 0$. There could be multiple causes for this shift away from GR, which we will explain below. Most likely, a combination of them are playing a role for GW230529.

Because the likelihood barely changes along the line of the correlation, small changes in the likelihood due to, for example, noise could push the posterior to certain values. Normally when having multiple detectors observing a signal, the expectation is that such fluctuations due to noise cancel each other out between the detectors since the noise is mostly uncorrelated. For single-detector events, the correlation between detectors cannot be used to prevent extra testing GR parameters being fitted to noise features. Because this event is single detector, the noise might indeed be playing a part in the shift of the posterior toward higher chirp mass and $\delta\hat{\varphi}_0$.

It is known that noise effects such as nonstationarity, non-Gaussianity, and glitches can impact tests of GR~\cite{Kwok:2021zny, LIGOScientific:2021sio, Gupta:2024gun, Maggio:2022hre} and even parameter estimation in GR~\cite{Chatziioannou:2019zvs, Cornish:2020dwh, Edy:2021par, Macas:2022afm}. More studies would be needed to see if biases like the shift away from GR at 0PN can be produced by noise and to see if biases in tests of GR due to noise effects are indeed enhanced for single-detector events. There was no transient noise found around GW230529~\cite{LIGOScientific:2024elc}, so we do not expect the noise to be majorly impacting our results.

The choice of prior could also significantly influence the posterior when there is such a degeneracy. For testing GR parameters that are strongly correlated with astrophysical parameters, this is a known issue, and it is therefore important to choose the correct priors~\cite{Payne:2023kwj}. We used a prior for the chirp mass that is uniform in the component masses, which translates to a preference for higher chirp-mass values. Since the waveform (and thus the likelihood) does not change much along this degeneracy, the posterior is likely shifted toward higher chirp mass and $\delta\hat{\varphi}_0$ because of this choice of prior.

\begin{figure}
    \centering
    \includegraphics[width=\columnwidth]{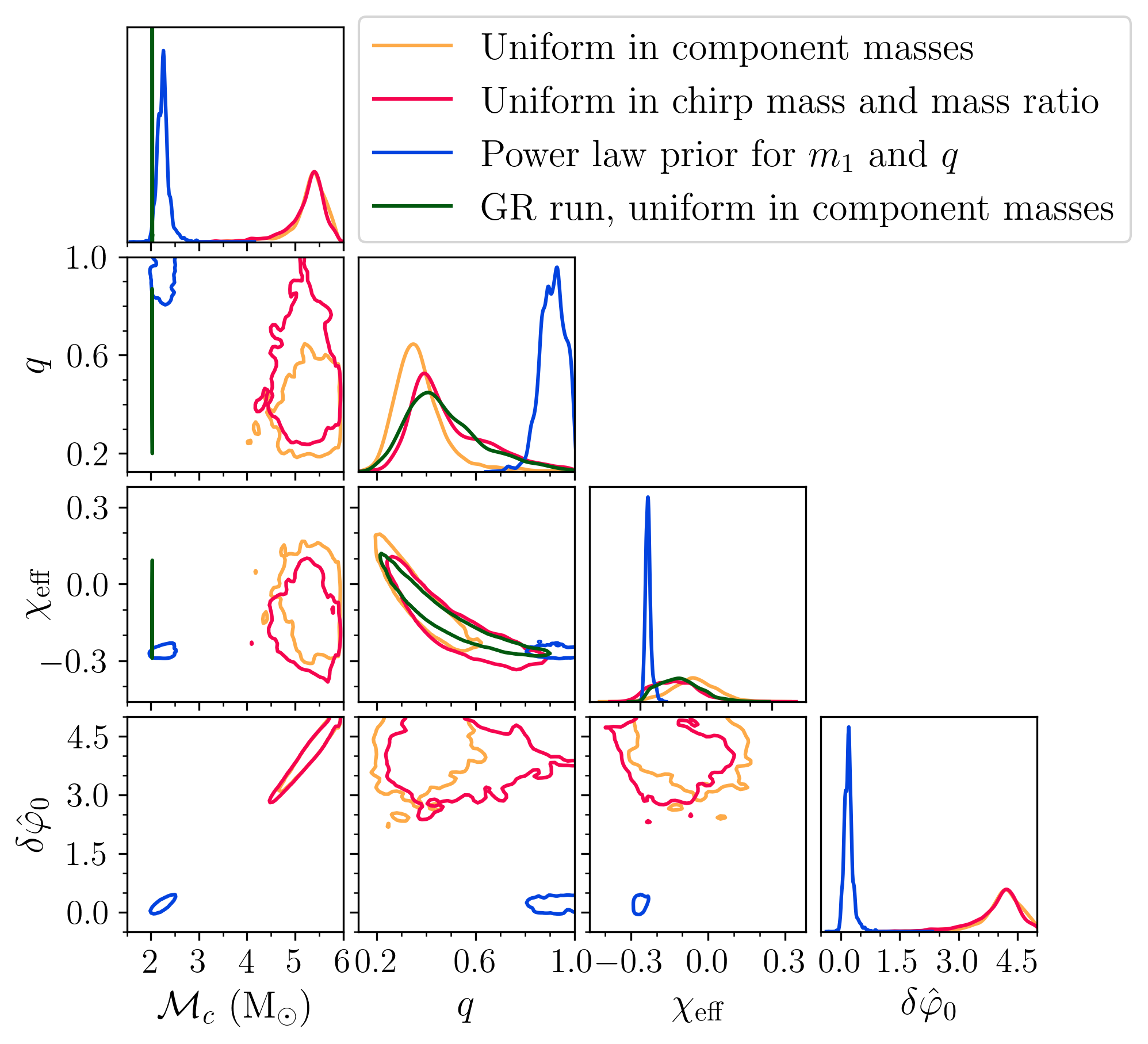}
    \caption{Results for the 0PN analysis using different priors for the masses obtained with \SEOBHM. We see that the low-mass-inclined power law prior (blue) gives results that are consistent with GR as opposed to the default uniform in component masses prior (orange) and the uniform in chirp mass and mass ratio prior (red). We also show the posteriors for a GR run using the default prior and \SEOBHM.}
    \label{fig:varying-prior}
\end{figure}

\citet{Payne:2023kwj} show that an astrophysically informed prior can help mitigate biases in tests of GR. To check the effect the choice of prior has on the 0PN results for GW230529, we also performed an analysis with a prior that resembles their astrophysical prior for the primary mass and mass ratio (see Eqs. (5) and (6) in Ref.~\cite{Payne:2023kwj}). Instead of inferring the powers $\alpha$ and $\beta$, we fix their values to $\alpha=3$ and $\beta=3$, which is consistent with their results for the 0PN test. Note that we extend their prior to lower masses, where the power law is no longer representative of the astrophysical population, but does exemplify a simple analytic prior that prefers lower masses and more equal masses. We also performed an analysis where we use a prior that is uniform in the chirp mass and mass ratio. The results are shown in Fig. \ref{fig:varying-prior}. We see that the low-mass-inclined power law prior (blue) gives results that are consistent with GR. The results also show more equal masses, which is expected considering the prior prefers this, and a negative effective spin. This shift toward more equal masses and negative effective spin also happens for GR when using certain population-informed priors~\cite{LIGOScientific:2024elc}. The results with the prior uniform in chirp mass and mass ratio (red) agree well with the results using our default prior that is uniform in the component masses (orange). This confirms that the apparent deviation from GR for $\delta\hat{\varphi}_0$ is prior dependent and most likely not a true deviation.

Another effect that could be causing the shift of the posterior towards higher chirp mass is some sampling issue. As can be seen in Fig.~\ref{fig:logl}, the likelihood is sharply peaked around the correlation line (i.e., it is extremely narrow and elongated). It is often difficult for the sampler to find such narrow features in the likelihood. We see in Fig.~\ref{fig:mismatch} that the low mismatch region becomes wider for high chirp mass, which means that the high-likelihood region also becomes wider for high chirp mass. This makes it easier for the sampler to find that part of the high-likelihood region. This could be part of the cause for the shift of the posterior toward high chirp mass and $\delta\hat{\varphi}_0$. Changing the parametrization and priors used could make it easier for the sampler to find the high likelihood region in the lower chirp mass regime.

Lastly, GW230529 being detected by only the LIGO Livingston detector might also lead to biases. For single-detector events, the extrinsic parameters are not well measured. Because of correlations, biases on extrinsic parameters could propagate to intrinsic ones, negatively affecting how well they are
measured compared to the case of an event detected by multiple detectors. The chirp mass, luminosity distance, and inclination are correlated through the amplitude of the signal. In GR, the chirp mass is being well measured in the phase and the amplitude is not needed to constrain it. Since the chirp mass is free to change together with $\delta\hat{\varphi}_0$ in the phasing for low-mass systems, the amplitude would help constrain the chirp mass. But for single-detector events, the inclination and distance are not well measured (multiple detectors are needed for that). Indeed, we also see that the luminosity distance is wider and shifted toward larger values compared to the GR run. This increase in correlations for single-detector events could be contributing to the bias in chirp mass and $\delta\hat{\varphi}_0$. This could also explain why this shift away from GR is not observed as strongly for the BNS GW170817~\cite{LIGOScientific:2018dkp} and the NSBH GW200115\_042309~\cite{LIGOScientific:2021sio}, which were both observed in three detectors.

\begin{figure*}
    \centering
    \includegraphics[width=\columnwidth]{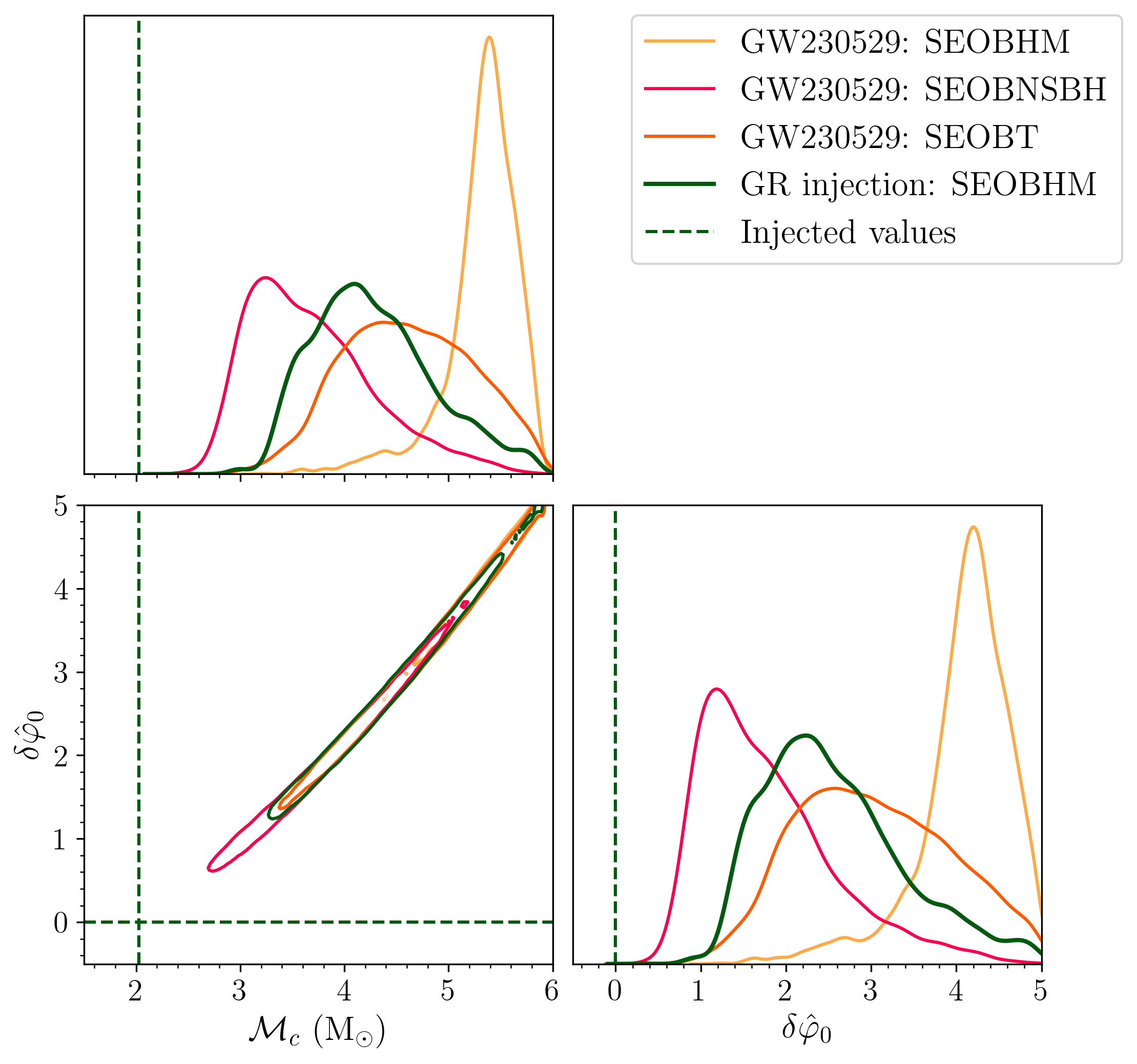}
    \includegraphics[width=\columnwidth]{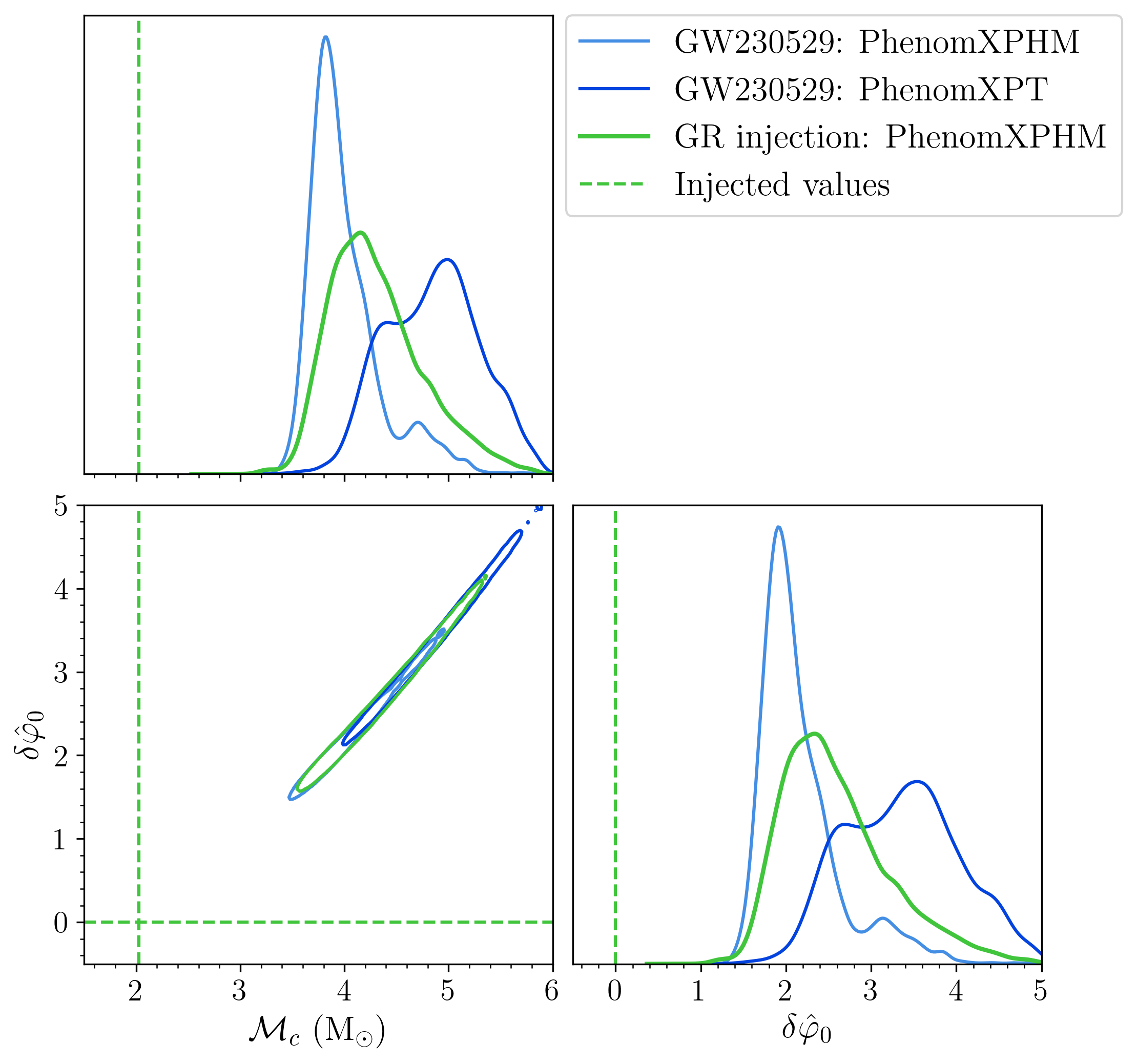}
    \caption{Posteriors for the chirp mass $\mathcal{M}_c$ and 0PN deviation parameter $\delta\hat{\varphi}_0$ for zero-noise injections of the maximum likelihood waveform from the corresponding GR run (green), compared to the posteriors obtained for GW230529 with the different waveform models. Left shows the posteriors obtained with FTI and right shows the results from TIGER. We see that the results from the GR injections are biased toward higher $\mathcal{M}_c$ and positive $\delta\hat{\varphi}_0$ and are similar to the GW230529 results.}
    \label{fig:dchi0-injection}
\end{figure*}

As a check to see that this shift away from GR is not a true deviation from GR, we performed parameter estimation on zero-noise injections of the maximum likelihood waveform from the corresponding GR run using \SEOBHM and \PhenomXPHM. Figure \ref{fig:dchi0-injection} shows that the posteriors obtained for the chirp mass and $\delta\hat{\varphi}_0$ (green) are again shifted toward higher values away from GR and the true chirp mass (green dashed lines), but they are different from the ones obtained for GW230529 using the same waveform models (light orange and light blue). The differences are of the same order as those we get between individual waveform models. From this we can conclude that the noise is most likely not the dominant factor in the shift away from GR, but we cannot fully exclude that it might be playing a part.

We conclude that for GW230529 the results show a strong correlation between the 0PN deviation parameter and the chirp mass, and that this correlation matches the expected degeneracy that would be there for a GR signal with such low masses. Even though the posterior excludes the GR values, we do not consider the 0PN results for GW230529 as evidence against consistency with GR, because the shift away from zero is along the line of this degeneracy. We also observe a similar shift away from GR for a zero-noise injection with the maximum likelihood GR waveform, which further confirms that the 0PN results for GW230529 are most likely not a true deviation from GR.

\subsection{Comparison with previous events}

\begin{figure*}
    \centering
    \includegraphics[width=\textwidth]{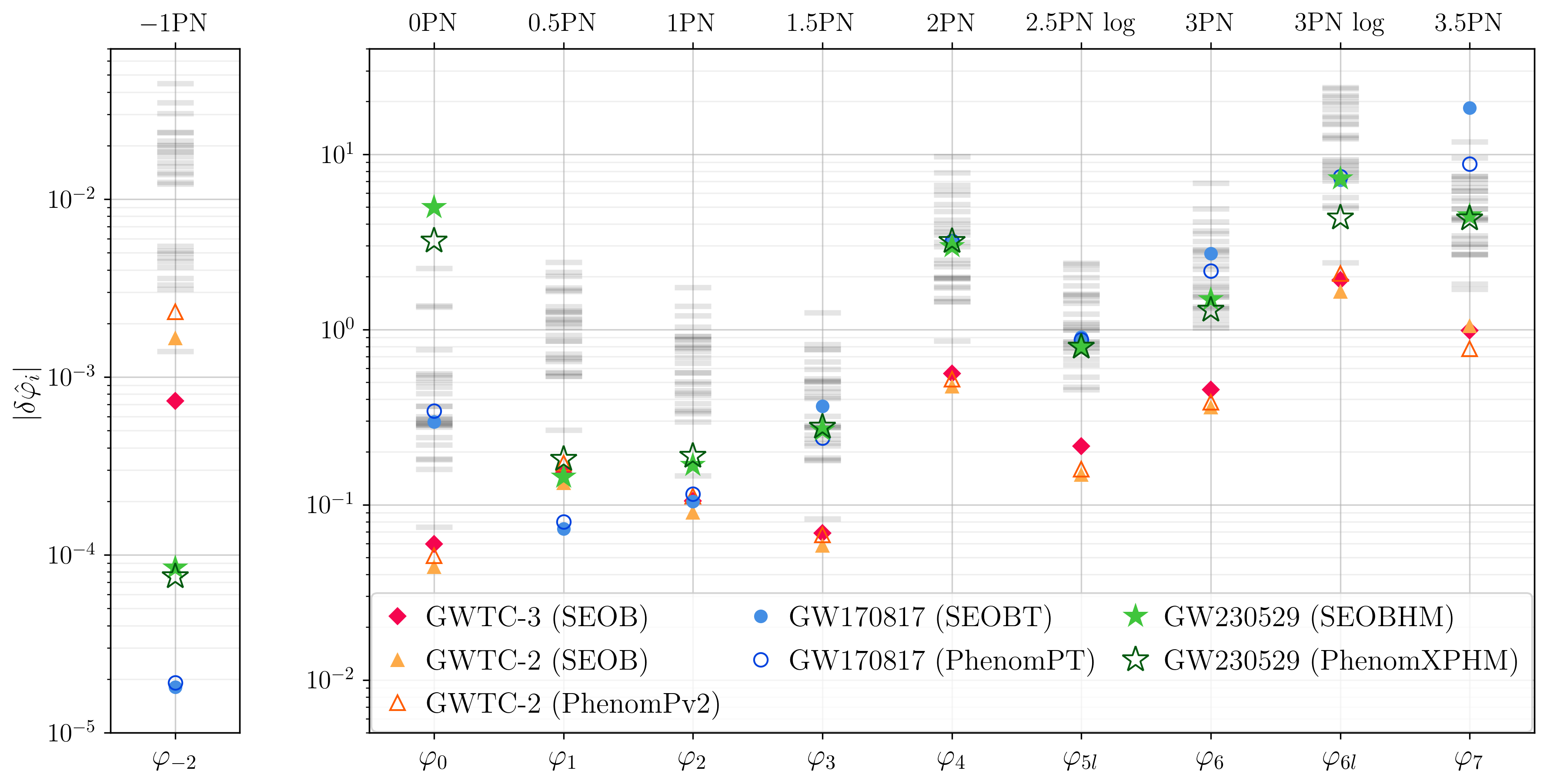}
    \caption{The bounds for the different deviation parameters $\delta\hat{\varphi}_n, \delta\hat{\varphi}_{n(l)}$ obtained with GW230529 (green stars) compared to previously obtained bounds. We show the results obtained with the BBH waveforms as a proxy for an NSBH with realistic tides, as discussed in Sec. \ref{sec:tides}. The orange triangles and red diamonds are bounds from GWTC-2 and GWTC-3, respectively, obtained by combining the posteriors for different events together (using \textsc{SEOBNRv4\_ROM} and \textsc{IMRPhenomPv2}). The blue circles are the bounds obtained with binary NS GW170817 (where they used \textsc{SEOBNRv4\_ROM\_NRTidalv2} and \textsc{IMRPhenomPv2\_NRTidalv2}). The grey stripes are bounds from individual events in GWTC-2 and GWTC-3. We see that GW230529 gives exceptionally tight bounds for the $-1$PN, 0.5PN, and 1PN deviation parameters. (Adapted from Ref.~\cite{LIGOScientific:2021sio}.)}
    \label{fig:bounds}
\end{figure*}

Figure \ref{fig:bounds} shows the 90\% upper bounds on the deviation coefficients found for GW230529 (green stars), as well as the combined bounds from testing GR with GWTC-2~\cite{LIGOScientific:2020tif} (orange triangles) and GWTC-3~\cite{LIGOScientific:2021sio} (red diamonds). The plot also shows for comparison the bounds from GW170817~\cite{LIGOScientific:2018dkp} (blue circles) and other bounds from individual events (grey stripes). Most noteworthy is the bound on the $-1$PN deviation of $|\delta\hat{\varphi}_{-2}| \lesssim 8 \times 10^{-5}$. This is a factor $\sim17$ better than previous bounds from the NSBH GW200115\_042309. It is also tighter than the combined constraint from GWTC-3 of $|\delta\hat{\varphi}_{-2}| \lesssim 8 \times 10^{-4}$, which combined the results from all NSBHs and BBHs in GWTC-3 for which the LVK performed tests of GR. The only previous $-1$PN bound that is tighter than our constraints from GW230529, is the results for BNS GW170817, from which a bound of $|\delta\hat{\varphi}_{-2}| \lesssim 2\times 10^{-5}$ was obtained. The bounds found at 0.5PN and 1PN are also relatively tight, with $|\delta\hat{\varphi}_1| \lesssim 0.2$ even being similar to the previous combined bounds.

We would expect the degeneracy at 0PN to also be there for other low-mass events. This is indeed the case for the BNS GW170817~\cite{LIGOScientific:2018dkp} and the NSBH GW200115\_042309~\cite{LIGOScientific:2021sio}. Those two events also show the shift away from GR toward positive $\delta\hat{\varphi}_0$, however, not as strong as for GW230529 and they still have support for GR. This is most likely due to those events being observed in three detectors, so there are fewer degeneracies with extrinsic parameters.

\section{Tests of ESGB gravity} \label{sec:esgb}

To illustrate the importance of the $-1$PN bounds from GW230529 for constraints on specific alternative theories of gravity, we have here a look at ESGB~\cite{Nojiri:2005vv, DeFelice:2010aj}. ESGB is a modified gravity theory where a scalar field is coupled to the Gauss-Bonnet density. It is described by the action
\begin{equation}
    S = \frac{1}{16\pi} \int \mathrm{d}x^4 \sqrt{-g} \left( R -2(\partial\phi)^2 + \ell_{\mathrm{GB}}^2 f(\phi) \mathcal{G}  \right). \label{eq:esgb-action}
\end{equation}
Here, $g$ is the metric determinant, $R$ is the Ricci scalar, $\phi$ is a scalar field with kinetic term $(\partial\phi)^2=g^{\mu\nu}\partial_\mu\phi\partial_\nu\phi$, and $\mathcal{G}$ is the Gauss-Bonnet invariant
\begin{equation}
    \mathcal{G} = R^{\mu\nu\rho\sigma} R_{\mu\nu\rho\sigma} - 4R^{\mu\nu} R_{\mu\nu} + R^2,
\end{equation}
where $R^{\mu\nu\rho\sigma}$ is the Riemann tensor and $R^{\mu\nu}$ the Ricci tensor.  The integral over a four-dimensional spacetime $\int d^4x\sqrt{-g}\, \mathcal{G}$ is a boundary term~\cite{Myers:1987yn}. The function $f(\phi)$, thus defined modulo a constant, specifies the theory. As for $\ell_{\mathrm{GB}}$, it is the Gauss-Bonnet coupling strength with dimension of length. We assume that matter fields are minimally coupled to $g_{\mu\nu}$.

In this paper, we consider the class of theories such that $f'(0)\neq 0$. Without loss of generality, the function $f(\phi)$ can be expanded as follows:
\begin{equation}
    f(\phi) = 2\phi+\mathcal O(\phi^2).\label{def:shiftSym}
\end{equation}
At leading order in $\phi$, the action \eqref{eq:esgb-action} reduces to that of shift-symmetric ESGB theories~\cite{Julie:2019sab}, which is invariant under the shift $\phi \rightarrow \phi + \Delta\phi$, where $\Delta\phi$ is a constant. This is equivalent to the small-$\phi$ approximation of Einstein-dilaton-Gauss-Bonnet for which $f(\phi)=e^{\gamma\phi}$ with a constant $\gamma$ (the value of $\gamma$ can be absorbed into the coupling constant). Here, we use the convention of Refs.~\cite{Julie:2019sab, Julie:2022huo,Julie:2022qux}. Another convention denotes the Gauss-Bonnet coupling by $\alpha_\mathrm{GB}$~\cite{Lyu:2022gdr}, which is related to our $\ell_{\mathrm{GB}}$ as $\alpha^2_\mathrm{GB} = \ell_{\mathrm{GB}}^4/(16\pi)$. 

It is possible to find the ESGB corrections to the GW phase during inspiral using the PN formalism. The leading-order correction to the frequency domain phase in ESGB appears at $-1$PN due to scalar dipole radiation. In the class of ESGB theories~\eqref{def:shiftSym} and at $\mathcal{O}\left(\ell_{\mathrm{GB}}^4\right)$, it reads~\cite{Lyu:2022gdr} % \comm{AB: Please add refs.} 
\begin{equation}
    \delta\hat{\varphi}_{-2} = -5\ell_{\mathrm{GB}}^4 \frac{\left(m_1^2 s_2 - m_2^2 s_1\right)^2}{168 m_1^4 m_2^4}, \label{eq:esgb-dchi-minus2}
\end{equation}
where $m_{1,2}$ are the masses of the compact objects, while $s_{1,2}$ were introduced in Refs.~\cite{Yunes:2016jcc,Lyu:2022gdr}. Here, $s$ is defined as the scalar monopole of a body at $\mathcal O(\ell_{\rm GB}^2)$, divided by that of a nonspinning BH with the same mass~\cite{Julie:2019sab}. The higher-order corrections have been fully computed up to 1PN (2PN relative order) for nonspinning binaries in the past~\cite{Yagi:2011xp, Sennett:2016klh, Julie:2019sab, Shiralilou:2020gah, Shiralilou:2021mfl, Julie:2022huo, Julie:2022qux}. These higher-order corrections are given in Appendix \ref{ap:esgb-corrections}. We also provide the full 1.5PN correction in Eq.~\eqref{eq:esgb-dchi3}, which is new to this paper.

In the class of ESGB theories \eqref{def:shiftSym}, NSs cannot carry scalar monopoles at $\mathcal O(\ell_{\rm GB}^2)$ so $s=0$~\cite{Yagi:2011xp,Yagi:2015oca,Lyu:2022gdr}, and for nonspinning BHs, $s=1$~\cite{Julie:2019sab}. So for a binary consisting of a NS and a nonspinning BH, Eq.~\eqref{eq:esgb-dchi-minus2} can be simplified to
\begin{equation}
    \delta\hat{\varphi}_{-2} = \frac{-5 \ell_{\mathrm{GB}}^4}{168 m_1^4}, \label{eq:esgb-nsbh}
\end{equation}
where $m_1$ is the mass of the BH. When computing the ESGB corrections for GW230529, we assume the secondary to be a NS independently of the waveform model used (remember that we use the BBH results as a proxy for the NSBH results with realistic tides).

\subsection{Constraints on ESGB by mapping agnostic results}

It is possible to get approximate constraints on the ESGB coupling from theory-agnostic tests. This is done by mapping the $-1$PN agnostic constraints to the ESGB coupling using the leading order correction. Inverting Eq.~\eqref{eq:esgb-nsbh} gives
\begin{equation}
    \ell_{\mathrm{GB}} = \left( \frac{-168 m_1^4 \delta\hat{\varphi}_{-2}}{5} \right)^{1/4}
\end{equation}
Note that we need to restrict ourselves to negative values of $\delta\hat{\varphi}_{-2}$. When mapping posteriors, we also need to reweight the samples to account for differing priors. To map from a prior uniform in $\delta\hat{\varphi}_{-2}$ to a prior uniform in $\ell_{\mathrm{GB}}$, the new sample weights $w$ become
\begin{equation}
    w = \frac{\partial \ell_{\mathrm{GB}}}{\partial (\delta\hat{\varphi}_{-2})} = \frac{\ell_{\mathrm{GB}}}{4\delta\hat{\varphi}_{-2}}.
\end{equation}

\begin{figure}
    \centering
    \includegraphics[width=\columnwidth]{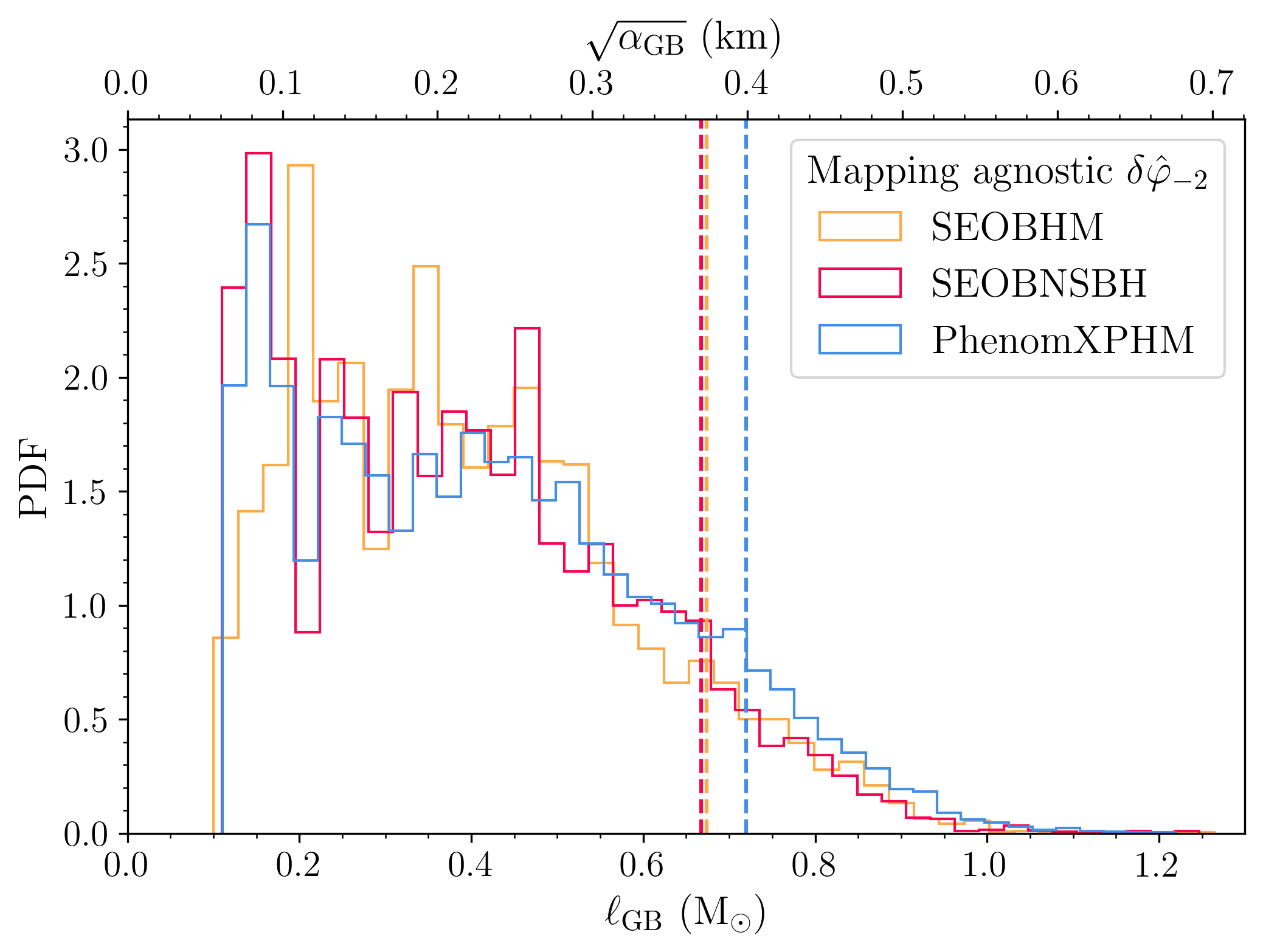}
    \includegraphics[width=\columnwidth]{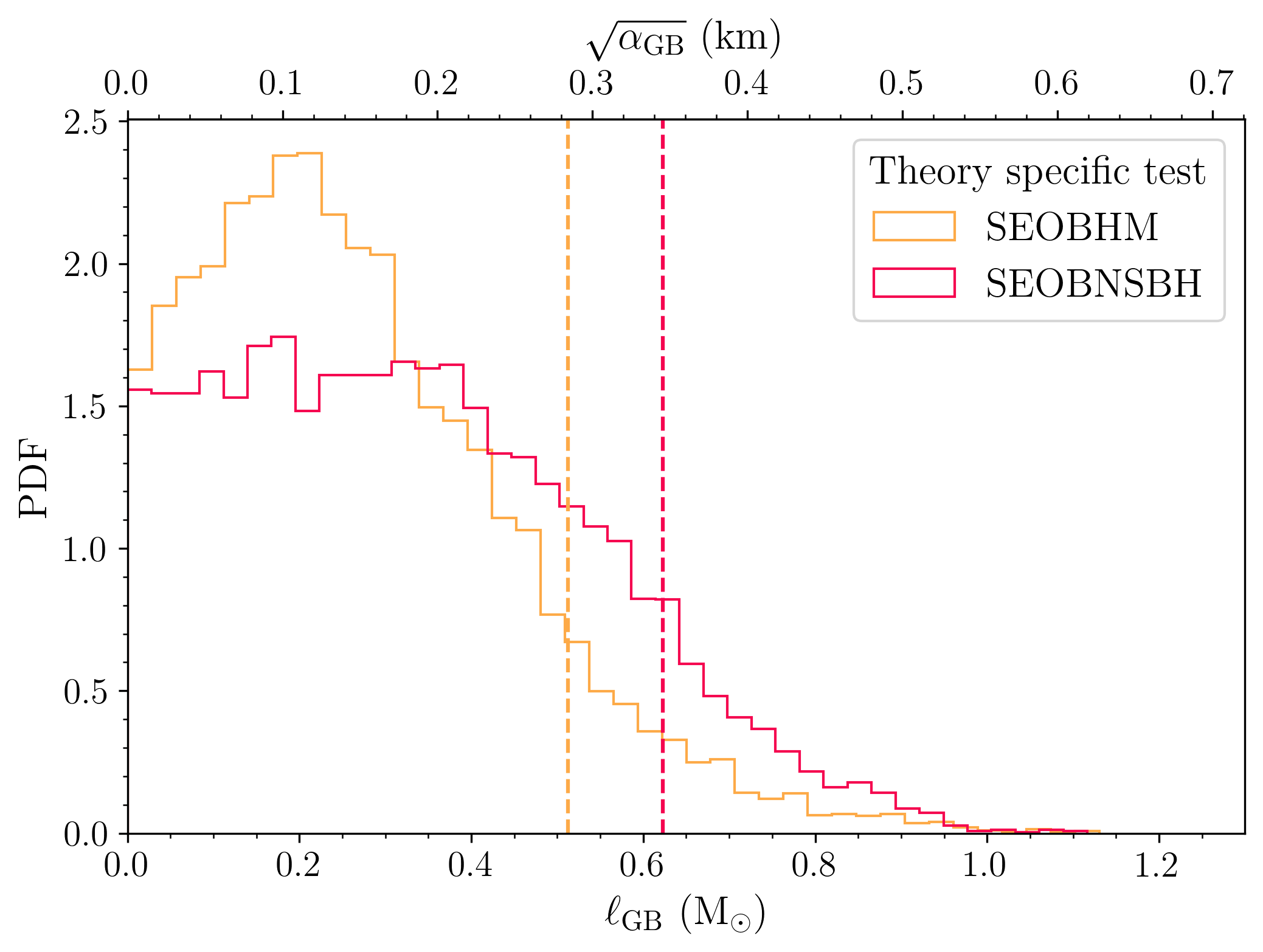}
    \caption{ Top: Posteriors for the ESGB coupling $\ell_{\mathrm{GB}}$ obtained by reweighting the $\delta\hat{\varphi}_{-2}$ posterior. Bottom: Posteriors for the Einstine-scalar-Gauss Bonnet coupling $\ell_{\mathrm{GB}}$ from the theory-specific test. The top axes show a slightly different definition of the ESGB coupling $\sqrt{\alpha_\mathrm{GB}} = {\ell_{\mathrm{GB}}}/{(2\pi^{1/4})}$ that is also commonly used.}
    \label{fig:esgb-posterior}
\end{figure}

The posteriors obtained by mapping the agnostic posteriors for $\delta\hat{\varphi}_{-2}$ to $\ell_{\mathrm{GB}}$ are shown in the top panel in Fig.~\ref{fig:esgb-posterior}. We see that the posteriors for results obtained with \SEOBHM (orange), \SEOBNSBH (red), and \PhenomXPHM (blue) do not significantly differ. The gap between $\ell_{\mathrm{GB}} \sim 0.1$ and 0 is not due to lack of support for GR. It is rather an effect of the sampling and reweighting due to the $\ell_{\mathrm{GB}}^4$ dependence of $\delta\hat{\varphi}_{-2}$. Values of $\ell_{\mathrm{GB}} \lesssim 0.1$ in this case require $\delta\varphi_{-2} \lesssim 10^{-8}$, which is four orders of magnitude smaller than the bound found. The limited number of samples is therefore causing a gap for small values. This is also the reason why the lower end of the posterior looks so ragged; there are only a few samples that have a relatively large weight. We would need many more samples to close this gap and make the posterior smooth for small $\ell_{\mathrm{GB}}$.

The 90\% upper bound on the ESGB coupling (indicated with dashed lines in Fig.~\ref{fig:esgb-posterior}) is $\ell_{\mathrm{GB}} \lesssim 0.67~\msun$ or $\sqrt{\alpha_\mathrm{GB}} \lesssim 0.37$~km (for \SEOBHM and \SEOBNSBH, for \PhenomXPHM the bound is $\ell_{\mathrm{GB}} \lesssim 0.72~\msun$ or $\sqrt{\alpha_\mathrm{GB}} \lesssim 0.40$~km). This bound can be improved by including higher PN order effects in ESGB. In the next Section, we include these effects in a theory-specific inspiral test for ESGB.

\subsection{Theory-specific test for ESGB}

Due to the flexibility of the FTI framework, we can use it not only to do a theory-agnostic inspiral test, but also for theory-specific inspiral tests~\cite{Mehta:2022pcn}. This is done by explicitly using the corrections to the frequency-domain phase from a specific theory. This means that we can not only use the leading-order corrections, but also the higher-order corrections. For ESGB, we can thus use all corrections up to 1.5PN, as given in Appendix \ref{ap:esgb-corrections}. We only add corrections to the dominant $(2,2)$-mode. We can then do Bayesian inference sampling over the ESGB coupling $\ell_{\mathrm{GB}}$, where we use a prior uniform in $\ell_{\mathrm{GB}}$. A posteriori, we check that the requirement $\left(\ell_{\mathrm{GB}}/m_1\right)^4 \ll 1$ holds, which comes from the small coupling approximation that is made in the derivation of the coefficients.

The results from the theory-specific ESGB test are shown in the bottom panel of Fig.~\ref{fig:esgb-posterior}.  Note that for this case the posterior looks much smoother for small $\ell_{\mathrm{GB}}$ than was the case for the mapping and there is no gap close to zero. That is because we are now sampling directly on $\ell_{\mathrm{GB}}$ so there is no problems with undersampling low values of the coupling.

The constraint on the ESGB coupling found using \SEOBHM (orange) is $\ell_{\mathrm{GB}} \lesssim 0.51~\msun$ or $\sqrt{\alpha_\mathrm{GB}} \lesssim 0.28$~km, and using \SEOBNSBH gives slightly worse constraints of $\ell_{\mathrm{GB}} \lesssim 0.62~\msun$ or $\sqrt{\alpha_\mathrm{GB}} \lesssim 0.35$~km (red). We notice that the bounds obtained with the theory-specific test are better than the bounds found from mapping the agnostic $-1$PN results. We note that in this work, the higher-order corrections to the phase for testing ESGB theory are incorporated only in the FTI framework and not in the waveforms for TIGER. This is because the implementation with FTI is more flexible, and we do not expect the results to change, as the $-1$PN mapping-based results are consistent.

These constraints on the ESGB coupling are better than any previously obtained constraints using GWs. From NSBHs, the best constraint so far was $\sqrt{\alpha_\mathrm{GB}} \lesssim 1.33$~km for GW200115\_042309~\cite{Lyu:2022gdr}. The best overall constraint was obtained with GW190814 and was $\sqrt{\alpha_\mathrm{GB}} \lesssim 0.37$~km (assuming it is a BBH; its secondary lies in the lower mass gap)~\cite{Lyu:2022gdr}. These bounds are also better than constraints obtained using low-mass x-ray binaries of $\sqrt{\alpha_\mathrm{GB}} \lesssim 1.9$~km~\cite{Yagi:2012gp} and NS EOS of $\sqrt{\alpha_\mathrm{GB}} \lesssim 1.29$~km~\cite{Saffer:2021gak}. Laboratory tests for ESGB in the weak field limit provide bounds that are at least 12 orders of magnitude weaker than strong-field tests~\cite{Elder:2022rak}.

\section{Conclusion}

In this work we have performed inspiral tests of GR with GW230529, the merger of two compact objects with masses $3.6_{-1.1}^{+0.8} M_\odot$ and $1.4_{-0.2}^{+0.6} M_\odot$. The tests used are parametrized inspiral tests that modify the phase in the frequency domain by allowing generic modifications of the PN coefficients.

We used two different versions of the parametrized inspiral test that employ different waveform approximants: FTI~\cite{LIGOScientific:2018dkp, Mehta:2022pcn} using the \textsc{SEOBNRv4} waveform family~\cite{Bohe:2016gbl, Cotesta:2018fcv, Cotesta:2020qhw, Dietrich:2019kaq, Matas:2020wab} and TIGER~\cite{Agathos:2013upa, Meidam:2017dgf} using the \textsc{IMRPhenomX} waveform family~\cite{Pratten:2020ceb, Garcia-Quiros:2020qpx, Colleoni:2023ple}. We compared the results of both tests. The various BBH waveform approximants gave consistent results, all of which are also consistent with GR, except for the 0PN deviation coefficient. The results are shown in Fig.~\ref{fig:violin-plot}.

At 0PN, there is a degeneracy between the chirp mass and the deviation parameter. This means that the chirp mass and the 0PN deviation parameter are free to move along the correlation given by Eq.~\eqref{eq:0pn-degeneracy} without significantly changing the waveform. For higher mass systems, this degeneracy is broken by higher PN contributions in the late inspiral and by the merger-ringdown. For GW230529, however, the merger-ringdown is outside the analyzed frequency band, so there is a strong correlation between $\mathcal{M}_c$ and $\delta\hat{\varphi}_0$ in the results, as can be seen in Fig. \ref{fig:dchi0-mc}. We notice that the posterior is shifted away from GR and toward higher chirp mass. Since this is also observed in a zero-noise GR injection, it is very likely a false deviation from GR. The cause of this shift is probably a combination of the choice of priors, sampling issues due to a sharply peaked likelihood, the event being single detector, and noise features.

Since the source of GW230529 is most likely an NSBH, we also performed analyses with waveforms that include tidal effects. The deviation-parameter posteriors obtained with these waveforms are broader and shifted away from GR compared to their BBH counterparts. This is due to correlations between the tidal deformability and the deviation parameters. For GW230529, the tidal deformabilities are not well constrained and even allow unrealistically high values~\cite{LIGOScientific:2024elc}, leading to the broad and shifted posteriors of the $\delta\hat{\varphi}_i$. By constraining the tides to more realistic values, the shift and broadening (mostly) disappear (see Fig. \ref{fig:tidal-effect}). We therefore use the results from the BBH waveform models when quoting bounds below.

The bounds obtained on the low-PN deviation parameters (except for 0PN) are some of the best to date. The bound obtained on dipole radiation ($-1$PN deviation) for GW230529 of $|\delta\hat{\varphi}_{-2}| \lesssim 8 \times 10^{-5}$ is an order of magnitude tighter than the combined bounds from GWTC-2~\cite{LIGOScientific:2020tif} and GWTC-3~\cite{LIGOScientific:2021sio}. The only tighter bound obtained with GWs is for the binary NS GW170817~\cite{LIGOScientific:2018dkp}. The bounds obtained at 0.5PN and 1PN are also relatively tight compared to previous constraints. Figure \ref{fig:bounds} compares the bounds for GW230529 with previous constraints from GWs.

The bounds on $-1$PN deviations allow us to constrain scalar-tensor theories that have their leading-order modification at $-1$PN. In particular, we have mapped the posteriors obtained for $\delta\hat{\varphi}_{-2}$ to the ESGB coupling $\ell_{\mathrm{GB}}$ using Eq.~\eqref{eq:esgb-dchi-minus2}. This gives us an upper bound on the ESGB coupling of $\ell_{\mathrm{GB}} \lesssim 0.67~\msun$. We have also performed a theory-specific test for ESGB gravity by implementing all corrections up to 1.5PN in the FTI framework. This allows us to sample directly over the ESGB coupling $\ell_{\mathrm{GB}}$ and obtain a slightly better bound of $\ell_{\mathrm{GB}} \lesssim 0.51~\msun$. Figure~\ref{fig:esgb-posterior} shows the posteriors obtained from both the mapping and the theory-specific test. These constraints on the ESGB coupling are the best obtained from GW signals to date.

In this work, we found no evidence for deviations from GR for GW230529. We obtained particularly tight constraints on deviations from GR at low-PN orders (with the exception of 0PN), demonstrating the importance of signals similar to GW230529 for tests of GR. We also discussed some challenges that arise when analyzing signals from low-mass systems such as GW230529, namely tidal effects and the degeneracy between the chirp mass and the 0PN deviation parameter. Future, more detailed studies of these effects will hopefully lead to ways to better account for, and perhaps even mitigate, their systematic biases.

When completing this paper, we became aware of an independent study~\cite{Gao:2024rel} which obtained constraints on Einstein-dilaton-Gauss-Bonnet gravity using GW230529. They obtain a bound of $\sqrt{\alpha_\mathrm{GB}} \lesssim 0.260$~km when including higher-order corrections and using \textsc{IMRPhenomXPHM}, which is similar to our bound of $\sqrt{\alpha_\mathrm{GB}} \lesssim 0.28$~km using the theory-specific test for ESGB with \textsc{SEOBNRv4HM\_ROM}.

\section*{Acknowledgments}

The authors would like to thank N.V. Krishnendu for reviewing the manuscript and providing useful feedback. We also thank Aditya Vijaykumar for his work on developing and maintaining the \textsc{Bilby TGR} package.
S.R. was supported by the research program of the Netherlands Organization for Scientific Research (NWO).
The authors are grateful for computational resources provided by the LIGO Laboratory and supported by National Science Foundation Grants PHY-0757058 and PHY-0823459. The authors are also grateful for the computational resources provided by Cardiff University supported by STFC grant ST/I006285/1, as well as computational resources from the Max Planck Institute for Gravitational Physics in Potsdam. 
This research has made use of data or software obtained from the Gravitational Wave Open Science Center (\href{https://www.gwosc.org}{https://www.gwosc.org}), a service of the LIGO Scientific Collaboration, the Virgo Collaboration, and KAGRA. This material is based upon work supported by NSF's LIGO Laboratory which is a major facility fully funded by the National Science Foundation, as well as the Science and Technology Facilities Council (STFC) of the United Kingdom, the Max-Planck-Society (MPS), and the State of Niedersachsen/Germany for support of the construction of Advanced LIGO and construction and operation of the GEO600 detector. Additional support for Advanced LIGO was provided by the Australian Research Council. Virgo is funded, through the European Gravitational Observatory (EGO), by the French Centre National de Recherche Scientifique (CNRS), the Italian Istituto Nazionale di Fisica Nucleare (INFN) and the Dutch Nikhef, with contributions by institutions from Belgium, Germany, Greece, Hungary, Ireland, Japan, Monaco, Poland, Portugal, Spain. KAGRA is supported by Ministry of Education, Culture, Sports, Science and Technology (MEXT), Japan Society for the Promotion of Science (JSPS) in Japan; National Research Foundation (NRF) and Ministry of Science and ICT (MSIT) in Korea; Academia Sinica (AS) and National Science and Technology Council (NSTC) in Taiwan.

\section*{Data availability}

The data that support the findings of this article are openly available~\cite{sanger_2026_19387685}.

\onecolumngrid

\appendix

\section{Priors used} \label{ap:priors}

The priors used in the analyses of GW230529 are given in the tables below. The names of the priors refer to the names in the \textsc{Bilby} package~\cite{Ashton:2018jfp}. The numbers indicate the range used. The priors are chosen to be agnostic, which means most of them are uniform in the parameter sampled or uniform in volume (e.g. for angles). The priors used for the deviation parameters and chirp masses are listed in Table \ref{tab:priors_dchi_mc}. The increase in width of the chirp-mass priors for low-PN deviations is because the chirp-mass posterior becomes broader due to correlations with the deviation parameters at low-PN order. The priors used for the spins are listed in Table \ref{tab:priors_spins} and differ between different waveform approximants. This is due to differing spin descriptions (aligned spins versus precessing spins) and due to limitations in the validity of the models. Lastly, all other priors are listed in Table \ref{tab:priors_other}.

\begin{table}
    \centering
    \begin{tabular}{lll}
        \hline
        \hline
        Deviation parameter & Prior & Chirp mass prior \\
        \hline
        $\varphi_{-2}$ & $\mathtt{Uniform}(-0.1, 0.1)$ & $\mathtt{UniformInComponentsChirpMass}(1.9, 2.2)$ \\
        $\varphi_{0}$ & $\mathtt{Uniform}(-5, 5)$\footnote{The prior was limited to $\mathtt{Uniform}(-0.5, 5)$ for \XPNRTidal due to problems with waveform wraparound like described in Appendix \ref{ap:wrap-around}.} & $\mathtt{UniformInComponentsChirpMass}(1.8, 6.0)$ \\
        $\varphi_{1}$ & $\mathtt{Uniform}(-2, 2)$ & $\mathtt{UniformInComponentsChirpMass}(1.9, 2.2)$ \\
        $\varphi_{2}$ & $\mathtt{Uniform}(-2, 2)$ & $\mathtt{UniformInComponentsChirpMass}(2.00, 2.05)$ \\
        $\varphi_{3}$ & $\mathtt{Uniform}(-2, 2)$ & $\mathtt{UniformInComponentsChirpMass}(2.00, 2.05)$ \\
        $\varphi_{4}$ & $\mathtt{Uniform}(-20, 20)$ & $\mathtt{UniformInComponentsChirpMass}(2.00, 2.05)$ \\
        $\varphi_{5l}$ & $\mathtt{Uniform}(-20, 20)$ & $\mathtt{UniformInComponentsChirpMass}(2.00, 2.05)$ \\
        $\varphi_{6}$ & $\mathtt{Uniform}(-20, 20)$ & $\mathtt{UniformInComponentsChirpMass}(2.00, 2.05)$ \\
        $\varphi_{6l}$ & $\mathtt{Uniform}(-30, 30)$ & $\mathtt{UniformInComponentsChirpMass}(2.00, 2.05)$ \\
        $\varphi_{7}$ & $\mathtt{Uniform}(-30, 30)$ & $\mathtt{UniformInComponentsChirpMass}(2.00, 2.05)$ \\
        \hline
        \hline
    \end{tabular}
    \caption{Priors used for the deviation parameter and the chirp mass for the different runs. The priors on the deviation parameter are uniform between the indicated range. The priors on the chirp mass are uniform in the component masses and between the indicated range.}
    \label{tab:priors_dchi_mc}
\end{table}

\begin{table}
    \centering
    \begin{tabular}{lll}
        \hline
        \hline
        Waveform approximant & Parameter & Prior \\
        \hline
        \multirow{2}*{\SEOBHM, \SEOB, \PhenomXHM} & $\chi_1$ & $\mathtt{AlignedSpin}(0, 0.99)$ \\
         & $\chi_2$ & $\mathtt{AlignedSpin}(0, 0.99)$ \\
        \hline
        \multirow{2}*{\SEOBNSBH} & $\chi_1$ & $\mathtt{AlignedSpin}(0, 0.9)$ \\
         & $\chi_2$ & $\mathtt{AlignedSpin}(0, 0.05)$ \\
        \hline
        \multirow{6}*{\PhenomXPHM} & $a_1$ & $\mathtt{Uniform}(0, 0.99)$ \\
         & $a_2$ & $\mathtt{Uniform}(0, 0.99)$ \\
         & $\theta_1$ & $\mathtt{Sine}(0, \pi)$ \\
         & $\theta_2$ & $\mathtt{Sine}(0, \pi)$ \\
         & $\phi_{12}$ & $\mathtt{Uniform}(0, 2\pi)$ \\
         & $\phi_{jl}$ & $\mathtt{Uniform}(0, 2\pi)$ \\
        \hline
        \multirow{6}*{\PhenomXP, \XPNRTidal} & $a_1$ & $\mathtt{Uniform}(0, 0.99)$ \\
         & $a_2$ & $\mathtt{Uniform}(0, 0.05)$ \\
         & $\theta_1$ & $\mathtt{Sine}(0, \pi)$ \\
         & $\theta_2$ & $\mathtt{Sine}(0, \pi)$ \\
         & $\phi_{12}$ & $\mathtt{Uniform}(0, 2\pi)$ \\
         & $\phi_{jl}$ & $\mathtt{Uniform}(0, 2\pi)$ \\
        \hline
        \hline
    \end{tabular}
    \caption{Priors on the spin parameters used for the different waveform approximants. The \texttt{AlignedSpin} prior gives the prior distribution of the aligned spin component based on the generic spin priors.}
    \label{tab:priors_spins}
\end{table}

\begin{table}
    \centering
    \begin{tabular}{ll}
        \hline
        \hline
        Parameter & Prior \\
        \hline
        $q$ & $\mathtt{UniformInComponentsMassRatio}(0.125, 1)$ \\
        $d_L$ & $\mathtt{UniformSourceFrame}(1, 1000)$ \\
        $\delta$ & $\mathtt{Cosine}(-\pi/2, \pi/2)$ \\
        $\alpha$ & $\mathtt{Uniform}(0, 2\pi)$ \\
        $\theta_{jn}$ & $\mathtt{Sine}(0, \pi)$ \\
        $\psi$ & $\mathtt{Uniform}(0, \pi)$ \\
        $\phi_\mathrm{ref}$ & $\mathtt{Uniform}(0, 2\pi)$ \\
        $t_c$ & $\mathtt{Uniform}(1369419318.6460938, 1369419318.8460937)$ \\
        $\Lambda_{1,2}$  & $\mathtt{Uniform}(0, 5000)$ \\
        \hline
        \hline
    \end{tabular}
    \caption{Priors used for the other GR parameters for all runs and waveforms. The prior on the mass ratio is such that it is uniform in the component masses. The prior on the geocentric time is a window of 0.2 seconds around the merger time. The tidal deformabilities are set to zero for BBH waveforms, and $\Lambda_1=0$ for \SEOBNSBH.}
    \label{tab:priors_other}
\end{table}

\section{False violations of GR when using too wide priors} \label{ap:wrap-around}

Large values of the deviation parameters can lead to significant changes in the length of the waveform. For more extreme cases, this can lead to the waveform wrapping around due to the choice of segment length of the data analyzed. In the case of GW230529, this problem can arise for, e.g., the 0.5PN deviation coefficient, where, for extreme negative values of $\delta\hat{\varphi}_1$, the post-trigger duration becomes more than the 2 seconds used in the analysis. When using wider priors on the deviation parameters, this can lead to false deviations of GR. For example, the posteriors for the 0.5PN deviation parameter would be peaking around $\delta\hat{\varphi}_1 \sim -9$. Normally, these unphysical waveforms that lead to wraparound would be penalized enough by having a low likelihood so as not to get any support in the posterior. Because GW230529 is a single-detector event, we suspect it instead picked up on some noise feature or artifact at the edge of the data segment analyzed, which lead to a high likelihood. When increasing the post-trigger duration to include a longer segment of data after the signal ends, this false violation of GR disappears. For the results presented in this work, we instead restricted the priors on the deviation parameters to avoid the unphysical waveforms that lead to wraparound.

\section{Higher-order ESGB corrections in the GW phase} \label{ap:esgb-corrections}

The corrections to the frequency-domain phase in ESGB can be computed using the PN formalism. The leading order corrections appear at $-1$PN (compared to GR), and the phase corrections have been fully computed up to 1PN (2PN relative order) for nonspinning binaries~\cite{ Sennett:2016klh, Julie:2019sab, Shiralilou:2020gah, Shiralilou:2021mfl, Lyu:2022gdr, Julie:2022huo}. We calculated as well the complete 1.5PN corrections \eqref{eq:esgb-dchi3}, which are new to this paper, using the recent fluxes at relative 2.5PN order of Ref.~\cite{Bernard:2022noq}. In the formalism described in Sec.~\ref{sec:model}, the corrections are given by
\begin{subequations}
\begin{align}
    \delta\hat{\varphi}_{-2} & = -5\ell_{\mathrm{GB}}^4 \frac{\left(m_1^2 s_2 - m_2^2 s_1\right)^2}{168 m_1^4 m_2^4}, \\
    \delta\hat{\varphi}_{0} & = -5\ell_{\mathrm{GB}}^4 \frac{659 m_1^4 s_2^2 + 1370 m_1^2 m_2^2 s_1 s_2 + 659 m_2^4 s_1^2 + 728\eta \left(m_1^2 s_2 - m_2^2 s_1\right)^2}{16128 m_1^4 m_2^4}, \\
    \delta\hat{\varphi}_{1} & = 25\pi \ell_{\mathrm{GB}}^4 \frac{\left(m_1^2 s_2 - m_2^2 s_1\right)^2}{56 m_1^4m_2^4}, \\
    \begin{split}
        \delta\hat{\varphi}_{2} & = \ell_{\mathrm{GB}}^4 \left \{ \frac{5 m_1^4 s_2^2 \left[-13792267 + 5588352 \delta - 17640 \eta (743 + 594 \eta)\right]}{290304 m_1^4 m_2^4(743 + 924 \eta)}  - \frac{5 m_2^4 s_1^2 \left[13792267 + 5588352 \delta + 17640 \eta (743 + 594 \eta)\right]}{290304 m_1^4 m_2^4(743 + 924 \eta)} \right.  \\ 
        & + \left. \frac{2 m_1^2 m_2^2 s_1 s_2 \left[56018615 + 3528 \eta (12239 + 14850 \eta)\right]}{290304 m_1^4 m_2^4(743 + 924 \eta)} \right\},
    \end{split} \\
    \delta\hat{\varphi}_{3} & = \ell_{\mathrm{GB}}^4 \frac{m_1^4 s_2^2 (-14363 + 1792 \delta - 4564 \eta) + 2 m_1^2 m_2^2 s_1 s_2 (-3557 + 4564 \eta) - m_2^4 s_1^2 (14363 + 1792 \delta + 4564 \eta)}{43008 m_1^4 m_2^4}. \label{eq:esgb-dchi3}
\end{align}
\end{subequations}
Here, $\ell_{\mathrm{GB}}$ is the ESGB coupling, $m_1, m_2$ are the masses of the two objects that also appear in the combinations $\eta = m_1 m_2 / (m_1 + m_2)^2$ and $\delta = (m_1 - m_2)/(m_1 + m_2)$. For NSs, $s=0$~\cite{Lyu:2022gdr} and for nonspinning BHs, $s=1$~\cite{Julie:2019sab, Julie:2022huo}.

The ESGB corrections to the phase above are complete up to 1.5PN for nonspinning binaries. By contrast, Ref.~\cite{Gao:2024rel} uses the corrections presented in Ref.~\cite{Lyu:2022gdr}, which are complete up to 1PN only. Indeed, at 1.5PN, the corrections of Ref.~\cite{Lyu:2022gdr} depend on the free parameter $f^{\rm GB}_3$ to represent the contributions from the fluxes at relative 2.5PN order, which were unknown at the time. Similarly, the 2PN corrections of Ref.~\cite{Lyu:2022gdr} depend on $f^{\rm GB}_4$ to represent contributions from the binding energy at 3PN, and from the (still unknown) fluxes at relative 3PN order. Note that in practice, both $f^{\rm GB}_3$ and $f^{\rm GB}_4$ are set to zero in Ref.~\cite{Lyu:2022gdr}.

\twocolumngrid

\bibliography{biblio}

\end{document}